\documentclass[aps,prx,preprint,onecolumn,citeautoscript,nofootinbib]{revtex4-1}

\usepackage{amsmath}
\usepackage{graphicx}
\usepackage{dcolumn}
\usepackage{bbm}
\usepackage{bm}
\usepackage{subfigure}
\usepackage{amssymb}
\usepackage{hyperref}
[    linkcolor=black,urlcolor=black]
\hypersetup{
    colorlinks=true,
    citecolor=blue,
    }
\usepackage{xcolor}
\usepackage{subfigure}
\usepackage{makecell}
\usepackage{mathtools}
\usepackage{booktabs}
\usepackage{xparse}
\usepackage{tikz-feynman}
\usepackage{color}
\usetikzlibrary{decorations.pathreplacing,decorations.markings,shapes.geometric}
\usepackage[papersize={8.5in,11in}]{geometry}
\usepackage{color}
\definecolor{darkblue}{rgb}{0.,0.,0.4}
\definecolor{darkred}{rgb}{0.5,0.,0.}
\definecolor{BlueViolet}{RGB}{138,43,226}
\definecolor{SkyBlue}{RGB}{30,144,255}
\definecolor{DarkGreen}{RGB}{0,100,0}

\newcommand{\eq}[1]{Eq. \eqref{#1}}
\newcommand{\bqa}{\begin{eqnarray}} 
\newcommand{\eqa}{\end{eqnarray}}
\newcommand{\nn}{\nonumber \\}

\stepcounter{secnumdepth}
\stepcounter{tocdepth}

\tikzset{
  wavy/.style={decorate,
  decoration={snake,amplitude=.4mm,segment length=1mm, post length=0mm,pre length=0mm}},
  on each segment/.style={
    decorate,
    decoration={
      show path construction,
      moveto code={},
      lineto code={
        \path [#1]
        (\tikzinputsegmentfirst) -- (\tikzinputsegmentlast);
      },
      curveto code={
        \path [#1] (\tikzinputsegmentfirst)
        .. controls
        (\tikzinputsegmentsupporta) and (\tikzinputsegmentsupportb)
        ..
        (\tikzinputsegmentlast);
      },
      closepath code={
        \path [#1]
        (\tikzinputsegmentfirst) -- (\tikzinputsegmentlast);
    },
   },
  },
  mid arrow/.style={postaction={decorate,decoration={
        markings,
        mark=at position .55 with {\arrow[#1]{stealth}}
      }}},
}

\begin{document}

\title{
Fermi liquids beyond the forward scattering limit:
 the role of non-forward scatterings for 
 scale invariance and instabilities
}
\author{
Han Ma$^1$
and
Sung-Sik Lee$^{1,2}$
}
\affiliation{ $^{1}$ Perimeter Institute for Theoretical Physics, Waterloo, Ontario N2L 2Y5, Canada}
\affiliation{
$^{2}$ Department of Physics $\&$ Astronomy, McMaster University,
1280 Main St. W., Hamilton, Ontario L8S 4M1, Canada }

\begin{abstract}

Landau Fermi liquid theory is a fixed point theory of metals 
that includes the forward scattering amplitudes 
 as exact marginal couplings.
However, the fixed point theory
that only includes the strict forward scatterings
is non-local in real space.
In this paper, we revisit the Fermi liquid theory 
using the field-theoretic functional renormalization group formalism
and show how the scale invariant fixed point emerges
as a local theory,
 which includes
 not only the forward scatterings 
 but also non-forward scatterings
 with small but non-zero momentum transfers.
In the low-energy limit,
the non-forward scattering amplitude
takes a scale invariant form.
If the bare coupling is attractive 
beyond a critical strength,
the coupling function exhibits a run-away flow drived by non-forward scattering amplitudes, 
signifying potential instabilities in particle-hole channels.
The pairing interaction also obeys a scaling relation 
if the center of mass momentum of Cooper pairs is comparable with energy. 
The coupling functions fully capture the universal low-energy dynamics of the collective modes 
and instabilities of Fermi liquids.
The divergence 
 of the cocupling function in the particle-hole channel beyond a critical interaction suggests an instability toward an ordered phase with a momentum that depends on the interaction strength.
At the critical interaction, the instability corresponds to
 the uniform Pomeranchuk or Stoner instability, but the momentum 
 of the leading instability becomes non-zero for stronger attractive interaction. 
In the particle-particle channel, 
the coupling function reveals the dynamics of the unstable mode associated with the BCS instability.
When an unstable normal metal evolves into the superconducting state, there exists a period in which 
 a superconducting state with 
 spatially non-uniform phase appears due to the presence of unstable Cooperon modes with non-zero momenta.

\end{abstract}

\maketitle

\section{Introduction}

As one of the most prevalent phases of matter, 
metals and their phase transitions contain rich physics 
that is central to our understanding of quantum materials.
Thus, Landau Fermi liquid theory of metals has been one of the main pillars of modern condensed matter physics\cite{LANDAU, landau1959theory,abrikosov2012methods,pines2018microscopic}. 
Introduced initially as a phenomenological model,
it only keeps the strict 
forward scattering amplitudes 
as interactions between quasiparticles.
Despite its immediate success as a phenomenological theory, it took more than thirty years 
to theoretically justify the validity of the theory\cite{abrikosov1958theory,nozieres1962derivation,luttinger1962derivation,
SHANKAR,
POLCHINSKI1,
PhysRevB.42.9967}.
From the renormalization group point of view,
 Landau Fermi liquid theory represents a low-energy fixed point.
Being a fixed point theory valid strictly at zero energy, 
it is rightly non-local in real space
at any finite length scale.
On the other hand, it is desirable to have an effective field theory of Fermi liquids valid 
below a small but non-zero energy scale.
Such an effective field theory must be local at length scales larger than the inverse of the energy scale.
It will allow one to use the powerful machinery of local field theory in  describing the emergence of Fermi liquids and their instabilities 
from a mid-infrared energy scale  down to the zero-energy limit.
The complete understanding of Fermi liquids beyond the zero energy limit is crucial for extracting scaling behaviours of physical observables at finite energies\cite{
PhysRevB.73.045128,
PhysRevB.69.121102,
2012LitJP..52..142P,
DASSARMA2021168495}.
It will also serve as a  solid reference point 
for theories of non-Fermi liquids\cite{
	HOLSTEIN,
	HERTZ,
	PLEE1,
	REIZER,
	PLEE2,
    VARMALI,
	MILLIS,
	ALTSHULER,
	YBKIM,
	NAYAK,
	POLCHINSKI2,
	ABANOV1,
	ABANOV3,
	ABANOV2,
	LOH,
	SENTHIL,
	SSLEE,
	MROSS,
	MAX0,
	MAX2,
	HARTNOLL,
	ABRAHAMS,
	JIANG,
	FITZPATRICK,
	DENNIS,
	STRACK2,
	SHOUVIK2,
	PATEL2,
	SHOUVIK,
	RIDGWAY,
	HOLDER,
	PATEL,
	VARMA2,
	EBERLEIN,
	SCHATTNER,
	SHOUVIK3,
	CHOWDHURY,
	VARMA3,
 PhysRevLett.128.106402,
SENTHIL,
SUNGSIKREVIEW,
PhysRevX.11.021005,
2022arXiv220407585D}.
For a recent progress toward this goal 
that uses bosonization,
see Refs.~
\cite{PhysRevB.49.10877,doi:10.1080/000187300243363,2022arXiv220305004D}.

\begin{figure}
    \centering
    \includegraphics[width=.4\textwidth]{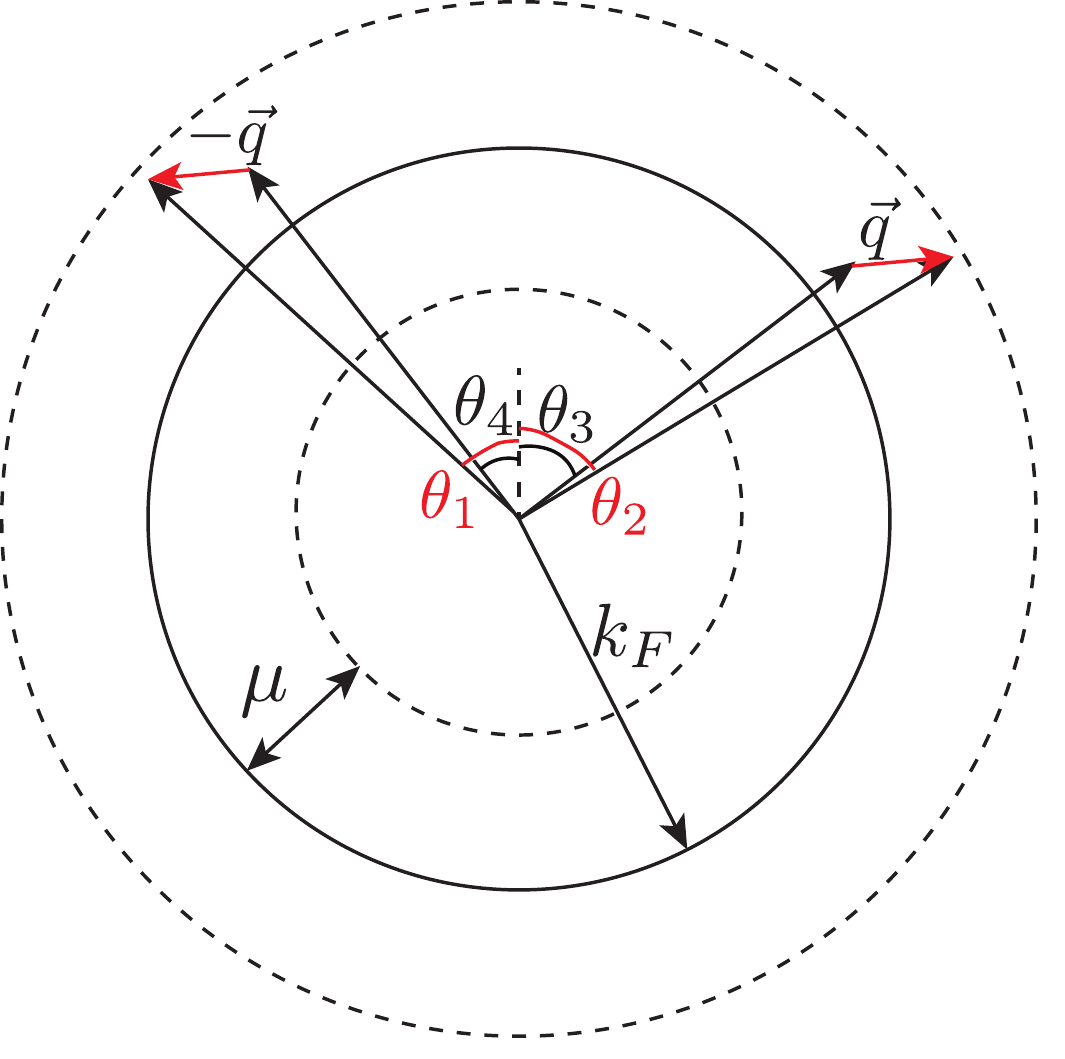}
    \caption{
    At energy scale $\mu$,
    two fermions within the energy shell of thickness $\mu$ can undergo non-forward scatterings by exchanging small but non-zero momentum $\vec q \sim \mu$.
    General scattering processes are captured by coupling functions that depend on four angles.
   Alternatively, it can be viewed as a function of two angles for the incoming fermions and momentum transfer $\vec q$.
   At low energies, the coupling function acquires 
   a non-trivial dependence on $\vec q/\mu$.
    }
    \label{fig:4-forward_scattering}
\end{figure}

In this work, we use the field-theoretic functional renormalization group scheme
to describe Landau Fermi liquid 
and its instabilities within the frame work of renormalizable  local effective field theory\cite{BORGES2023169221}. 
The key ingredient of our work is the non-forward scatterings.
A local effective field theory 
 must include non-forward scatterings 
because, at any non-zero energy scale, 
fermions can exchange non-zero momenta
while staying close to the Fermi surface 
within a thin energy shell.
Let $\lambda^{\theta_1 \theta_2}_{\theta_4 \theta_3}$
represent the coupling function
that describes the scattering 
of two low-energy fermions
from angles $(\theta_4,\theta_3)$
to $(\theta_1,\theta_2)$
(Fig.~\ref{fig:4-forward_scattering}).
While Landau Fermi liquid theory 
 only includes the strict forward scattering amplitude
($\lambda^{\theta_1 \theta_2}_{\theta_1 \theta_2}$), 
the full coupling function 
does depend on 
$\theta_1-\theta_4$ and 
$\theta_2-\theta_3$
non-trivially.
At a non-zero energy scale $\mu$,
the coupling function changes
smoothly but significantly 
as the differences in angles 
change by $\mu/k_F$,
where $k_F$ is the Fermi momentum.
If a UV theory is within the basin of attraction of the Fermi liquid fixed point,
the coupling function flows to a scale invariant form 
in the low-energy limit.
The scale invariance becomes manifest once the transferred momentum is scaled along with the energy scale.

While the strict forward scattering amplitude is exactly marginal\cite{SHANKAR}, 
the non-forward scattering amplitude does receive quantum corrections.
The non-trivial RG flow of the general quartic coupling function can drive
instabilities in particle-hole channels.
If the UV theory has a sufficiently strong attractive interaction,
the coupling function exhibits a run-away 
renormalization group (RG) flow to the strong coupling
region due to the non-trivial renormalization of the non-forward scatterings.
This implies that the general coupling function is a part of the low-energy data that should be kept within 
the low-energy effective theory with a small but non-zero energy cutoff. 
Within the local effective field theory, one should also consider general scattering processes 
in the particle-particle channel
by including interactions of Cooper pairs with small but non-zero center of mass momenta.
A scaling relation emerges 
 in the general pairing interaction 
once the center of mass momentum is scaled along with the energy scale.

The rest of the paper is organized in the following way.
In Sec.~\ref{sec:eff_theory}, we introduce the local effective field theory for Fermi liquids.
Sec.~\ref{sec:ZS} discusses the effect of non-forward scattering in the near forward scattering channels.
In Sec. \ref{sec:FL}, we present the scale invariance coupling function that emerges at Fermi liquid fixed points.
Sec.~\ref{sec:instability} discusses the instability toward symmetry broken states
driven by the flow of non-forward scattering amplitudes.
The scaling behaviour of the general pairing interaction is discussed in 
Sec.~\ref{sec:BCS}.
In Sec. \ref{sec:experiment},
we discuss the experimental implications of the momentum-dependent coupling functions.
We show that the coupling functions fully encode the universal low-energy dynamics of both stable and unstable collective modes of Fermi liquids.
Besides reproducing known results on the zero sound mode, we make predictions on instabilities in both particle-hole and particle-particle channels.
Finally, Sec.~\ref{sec:summary} summarizes the work.

\section{Local effective field theory \label{sec:eff_theory}}

We consider a circular Fermi surface of spinful fermions that are subject to short-range interactions in two space dimensions.
The following discussion can be generalized to 
higher dimensions in a straightforward way.
The partition function is written as 
$Z=\int \mathcal{D}\psi^\dag \mathcal{D}\psi ~e^{-S}$
and the action reads
\begin{eqnarray}
S &=&
\sum_{\sigma =\pm }
\int \frac{d\mathcal{V}_2}{(2\pi)^3}
~
\psi^\dag_{\sigma}(\omega,\vec{k}) 
(-i \omega +\varepsilon_{\vec{k}})
\psi_{\sigma}(\omega,\vec{k})  +
\frac{1}{4} 
\int \frac{d\mathcal{V}_4}{(2\pi)^9} 
~\sum_{\sigma_{1,2,3,4}=\pm  }
(\lambda)_{\vec{k}-\frac{\vec{q}}{2},\sigma_4;\vec{p}+\frac{\vec{q}}{2},\sigma_3}^{\vec{k}+\frac{\vec{q}}{2},\sigma_1;\vec{p}-\frac{\vec{q}}{2},\sigma_2} 
\nonumber\\
&\times&~
\psi^\dag_{\sigma_1}(\Omega+\frac{\omega}{2},\vec{k}+\frac{\vec{q}}{2})
\psi^\dag_{\sigma_2}(\Omega'-\frac{\omega}{2},\vec{p}-\frac{\vec{q}}{2})
\psi_{\sigma_3 }(\Omega'+\frac{\omega}{2},\vec{p}+\frac{\vec{q}}{2})
\psi_{\sigma_4}(\Omega-\frac{\omega}{2},\vec{k}-\frac{\vec{q}}{2}),
\label{eq:action}
\end{eqnarray}
where $d \mathcal{V}_2=d\omega d^2 \vec{k}$ and $d\mathcal{V}_4 = d\omega d\Omega d\Omega'  d^{2} \vec{k} ~d^{2} \vec{p} ~d^{2} \vec{q}$ are the integral measures of the quadratic and quartic terms.
$\psi_{\sigma}(\omega,\vec{k})$ 
denotes the fermionic field of spin $\sigma$,
momentum $\vec{k}$ and frequency $\omega$. 
The bare dispersion is written as
$\varepsilon_{\vec{k}}=\frac{1}{2m} (k^2- k_F^2) $,
where $k_F$ is the Fermi momentum.
$\lambda$ is the four-fermion coupling,
which is a function of momenta of the incoming and outgoing fermions.

At low energies, we focus on 
fermions that are close to the Fermi surface.
In defining the low-energy scaling limit of the theory,
it is convenient to use the polar coordinate, where
the two-dimensional momentum of fermion is written as 
$\vec{k}=(k_F+\kappa) \left( \cos \theta, \sin \theta \right)$.
$\kappa$ denotes the deviation of 
$|\vec k|$ from $k_F$,
and $\theta$ is the polar angle.
Accordingly, the fermion field is written as 
$\psi_{\sigma;\theta}(\omega, \kappa) \equiv
\psi_{\sigma}(\omega, \vec k)$. 
Low-energy effective field theories 
for finite number of low-energy fields
are characterized by a small number of coupling constants.
In metals, Fermi surfaces support infinitely many gapless modes 
 as the angle around the Fermi surface plays the role of a continuous flavor.
Accordingly, the coupling constants are promoted to coupling functions that depend on angles.
In the strict zero energy limit,
only two channels of interactions are allowed
by the momentum conservation. 
The first is the forward scatterings 
and the other is the pairing interactions.
Those interactions involve 
pairs of fermions 
with zero center of mass momentum in the particle-hole and particle-particle channels, respectively.
However, the interaction that only includes the strict forward scattering and 
the BCS interaction is non-local in the real space.
At small but non-zero energies, the locality forces us to include interactions in which 
 fermion pairs have small but non-zero center of mass momenta.
This leads to the local low-energy effective action, 
\begin{eqnarray}
S
&=& 
k_F 
\int \frac{d\bar{\mathcal{V}}_2 }{(2\pi)^3}
~\sum_{\sigma =\pm} \psi^\dag_{\sigma;\theta}(\omega, \kappa ) (-i \omega +v_F \kappa  )\psi_{\sigma;\theta}(\omega, \kappa) 
\nn &+&
\frac{k_F^2}{4}
 \int \frac{d\bar{\mathcal{V}}_4^{(0)}}{(2\pi)^9}  ~\sum_{\sigma_{1,2,3,4}=\pm}
(\lambda_0)^{\theta- \vartheta_\theta^S
, \sigma_1;
~ \theta'+\vartheta_{\theta'}^S, \sigma_2}_{
  \theta+ \vartheta_\theta^S, \sigma_4;
~ \theta'- \vartheta_{\theta'}^S, \sigma_3
 } \nonumber \\
&&\times
\psi^\dag_{\sigma_1;\theta- \vartheta_\theta^S}[\Omega+ \omega/2,\kappa + k_F \vartheta^C_\theta ]
\psi^\dag_{\sigma_2; \theta'+ \vartheta_{\theta'}^S}[\Omega'- \omega/2,\rho-k_F \vartheta^C_{\theta'} ]
\nonumber\\
&&\times \psi_{\sigma_3;\theta'- \vartheta_{\theta'}^S}[\Omega'+ \omega/2,\rho +k_F\vartheta^C_{\theta'}]
\psi_{\sigma_4; \theta+ \vartheta_{\theta}^S  }[\Omega- \omega/2,\kappa -k_F\vartheta^C_{\theta}] \nonumber \\
%%BCS
&+& 
\frac{k_F^2}{4} 
\int \frac{d\bar{\mathcal{V}}_4^{(1)}}{(2\pi)^9}  ~\sum_{\sigma_{1,2,3,4}=\pm}
(\lambda_1)^{\theta'- \varphi_{\theta'}^S
, \sigma_1;
~\theta'+ \varphi_{\theta'}^S +\pi , \sigma_2}_{
 \theta- \varphi_\theta^S, \sigma_4;
~ \theta+ \varphi_\theta^S +\pi, \sigma_3
 } \nonumber \\
&&\times
\psi^\dag_{\sigma_1; \theta'- \varphi_{\theta'}^S}[
\Omega'+ \omega/2,\rho + k_F \varphi_{\theta'}^C]
\psi^\dag_{\sigma_2;  \theta'+ \varphi_{\theta'}^S+\pi}[
-\Omega'+ \omega/2,\rho-k_F \varphi_{\theta'}^C ]\nonumber\\
&&\times \psi_{\sigma_3; \theta+ \varphi_\theta^S +\pi}[
-\Omega+\omega/2,\kappa -k_F \varphi_{\theta}^C]
\psi_{\sigma_4; \theta- \varphi_\theta^S  }[
\Omega+\omega/2,\kappa +k_F \varphi_{\theta}^C].
\end{eqnarray}
Here, 
$d\bar{\mathcal{V}}_2=d\omega d \kappa d \theta$ denotes the integral measure associated with 
a unit length along the Fermi surface. 
The quartic interaction has been divided into the forward and BCS channels whose integral measures are 
$d\bar{\mathcal{V}}_4^{(0)}=
d\omega d\Omega d\Omega'  
dq q 
d \kappa d \rho  
d \phi 
d\theta d\theta'
$ and $d\bar{\mathcal{V}}_4^{(1)}=
d\omega d\Omega d\Omega'  
dQ Q  d \kappa d \rho  
d \Phi d\theta d\theta'$, respectively.
Here $\lambda_0$  and $\lambda_1$ are the local coupling functions that include the forward and BCS scatterings\footnote{
It is noted that 
both $\lambda_0$, $\lambda_1$
includes the process in which two fermions with almost zero center of mass momentum go through a near forward scattering.
However, the phase space for such overlap is a set of measure zero in the low-energy limit.}.
They allow the total momenta 
of particle-hole and particle-particle pairs,
denoted as $\vec q$
and $\vec Q$ respectively,
to be non-zero.
$\vec q$  and $\vec Q$
are also written in the polar coordinate as
$\vec q=q(\cos \phi, \sin\phi)$,
$\vec Q=Q(\cos \Phi, \sin\Phi)$,
where these momenta
are not measured relative to Fermi momentum 
unlike the momenta of fermions.
In order for the fermions to stay close to the Fermi surface, 
the magnitudes of $\vec q$ and $\vec Q$ must to bounded by $\mu$  
 at energy scale $\mu$.
However, it is crucial to allow the momentum transfer to be flexible within that range in order to keep the locality of the effective theory.
$\vartheta^S_{\theta} = \frac{q}{2k_F} \sin(\theta-\phi)$
and
$\varphi^S_\theta = \frac{Q}{2k_F} \sin(\theta-\Phi)$ 
denote the deviation of the angles 
away from the
the strict forward and BCS scattering channels, respectively.
Similarly,
$\vartheta^C_{\theta} = \frac{q}{2k_F} \cos(\theta-\phi)$
and
$\varphi^C_\theta = \frac{Q}{2k_F} \cos(\theta-\Phi)$
determine the shift of the fermion energy caused by non-zero $q$ and $Q$ in the near forward and BCS scattering channels, respectively.
The theory is specified by two parameters,
$k_F$, $v_F$ and two coupling functions,
$\lambda_0$, $\lambda_1$.

Under the scale transformation, 
\begin{eqnarray}
{\omega}_s = \omega/s, \quad {\Omega}_s &=& \Omega/s, \quad {\Omega}_s' = \Omega'/s, \nonumber\\
{\kappa}_s = \kappa/s,  \quad
{\rho}_s &=& \rho/s, \quad
{q}_s = q/s, \quad 
{Q}_s = Q/s, 
\nonumber\\
{\psi}_{s; \sigma;\theta}({\omega}_s,  {{\kappa}_s}) &=& 
s^{2} \psi_{\sigma;\theta}(\omega, {\kappa}) 
\label{eq:scale_transformation}
\end{eqnarray}
that leaves angles unchanged,
the coupling constants ($v_F$, $k_F$) and coupling functions ($\lambda_0$, $\lambda_1$)
 are transformed as
\begin{eqnarray}
{k}_{s;F} =  k_F/s, && 
{v}_{s;F} =  v_F, \nonumber \\
\vartheta^{A}_{s;\theta}=\frac{q_s}{2k_{s;F}} \mathbb{A} (\theta-\phi) = \vartheta^{A}_{\theta}, && \varphi^{A}_{s;\theta} =\frac{Q_s}{2k_{s;F}}\mathbb{A} (\theta-\Phi) = \varphi^{A}_{\theta},\quad \textrm{ with } A=S,C \textrm{ and } \mathbb{A} = \sin, \cos \nonumber\\
(\lambda_{s;0})^{
 \theta-\vartheta^S_{\theta},\sigma_1;
  \theta'+ \vartheta^S_{\theta'},\sigma_2}_{
   \theta+\vartheta^S_{\theta},\sigma_4;
  \theta'-\vartheta^S_{\theta'},\sigma_3
 }  
 &=&
  s ( \lambda_0)^{
 \theta-\vartheta^S_{\theta},\sigma_1;
  \theta'+ \vartheta^S_{\theta'},\sigma_2}_{
   \theta+\vartheta^S_{\theta},\sigma_4;
  \theta'-\vartheta^S_{\theta'},\sigma_3
 } , \\ \nonumber
 %%BCS
 (\lambda_{s;1})^{
 \theta'-\varphi^S_{s,\theta'},\sigma_1;
  \theta' +\varphi^S_{s,\theta'}+\pi ,\sigma_2}_{
  \theta-\varphi^S_{s,\theta},\sigma_4;
  \theta+\varphi^S_{s,\theta}+\pi ,\sigma_3
 } & = &
 s (\lambda_1)^{
 \theta'-\varphi^S_{\theta'},\sigma_1;
  \theta' +\varphi^S_{\theta'}+\pi ,\sigma_2}_{
  \theta-\varphi^S_{\theta},\sigma_4;
  \theta+\varphi^S_{\theta}+\pi ,\sigma_3
 }. 
\end{eqnarray}
A few comments on the tree-level scaling is in order.
First, the four-fermion couplings have scaling dimension $-1$.
Nonetheless, the low-energy effective field theory must include them because they can give rise to infrared singularity.
In metals, the degree of IR singularity that a coupling can create does not necessarily match its scaling dimension because of 
the scale associated with the Fermi momentum.  
Consequently, the notion of renormalizable field theory 
needs to be generalized for metals\cite{BORGES2023169221}.
Second,  $k_F$ has scaling dimension $1$ and runs toward infinity in the low-energy  ($s \rightarrow 0$) limit.
This is because the size of the Fermi surface measured in the unit of the floating energy scale increases 
as the low-energy limit  is taken.
Here, $k_F$ plays the role of a metric that controls the `proper' size of Fermi surface\footnote{
Alternatively, one can adopt a scaling in which the angle is rescaled with a fixed $k_F$\cite{PhysRevB.78.085129}.}.
Third, the action has scaling dimension $1$ rather than $0$ under this scale transformation\footnote{
The action that includes higher order interactions also has dimension $1$ under the current scale transformation.
However, we don't include them here as they do not give rise to IR singularities\cite{BORGES2023169221}.
}.
This is an unusual choice in that action is regarded dimensionless in most field theories.
However, this is natural for theories with continuously many gapless modes where 
the manifold that supports gapless modes comes with a momentum scale.
The fact that the low-energy effective action has a positive dimension reflects the fact that 
the number of patches connected by non-forward scattering increases with decreasing energy\cite{PhysRevB.72.054531}.
At low energies,
the magnitude of typical momentum transfer $\vec q$ in the non-forward scattering decreases,
which makes the number of decoupled patches 
increases at low energies.

Now, we consider quantum corrections that renormalize the couplings.
We first define the couplings in terms of physical observables.
$k_F$ does not receive quantum corrections
due to the Luttinger's theorem.
$v_F$ and the coupling functions are defined through the two-point and the four-point vertex functions through
\bqa
\frac{\partial }{\partial \kappa} {\rm Re} \varGamma_{(2)}(k^*)
 = v_F + F_{1}, && 
 i \frac{\partial }{\partial \omega} {\rm Im} \varGamma_{(2)}(k^*)
 = 1+F_{2}, 
\label{eq:RG3}\\
 \left( \varGamma_{(4)} \right)^{
k_1^*, \sigma_1;
k_2^*, \sigma_2}_{
k_4^*, \sigma_4;
k_3^*, \sigma_3} 
 &=& 
(\lambda_0)^{
 \theta-\vartheta^S_{\theta},\sigma_1;
  \theta'+ \vartheta^S_{\theta'},\sigma_2}_{
   \theta+\vartheta^S_{\theta},\sigma_4;
  \theta'-\vartheta^S_{\theta'},\sigma_3
 } 
  +F_3, 
 \label{eq:RG4}
  \\
 %%BCS
\left( \varGamma_{(4)} \right)^{
p_1^*, \sigma_1;
p_2^*, \sigma_2}_{
p_4^*, \sigma_4;
p_3^*, \sigma_3} 
&=&
(\lambda_1)^{
 \theta'-\varphi^S_{\theta'},\sigma_1;
  \theta' +\varphi^S_{\theta'}+\pi ,\sigma_2}_{
  \theta-\varphi^S_{\theta},\sigma_4;
  \theta+\varphi^S_{\theta}+\pi ,\sigma_3
 }
 + F_4.
 \label{eq:RG5}
\eqa
Here, 
$\varGamma_{(2)}$ 
and
$\varGamma_{(4)}$ 
represent 
the two-point and four-point vertex functions, respectively.
The energy-momentum vectors used to impose the RG condition are chosen to be
$k^*  = $$ \Bigl( \mu, k_F \cos \theta, k_F \sin \theta \Bigr)$, 
$k_1^* =$$ \Bigl( 3\mu, k_F \cos \theta + \frac{q}{2} \cos \phi, k_F \sin \theta + \frac{q}{2} \sin \phi \Bigr)$, 
$k_2^* = $$\Bigl( -\mu, k_F \cos \theta' - \frac{q}{2} \cos \phi, k_F \sin \theta' - \frac{q}{2} \sin \phi \Bigr)$, 
$k_3^* = $$\Bigl( \mu, k_F \cos \theta' + \frac{q}{2} \cos \phi, k_F \sin \theta' + \frac{q}{2} \sin \phi \Bigr)$, 
$k_4^* = $$\Bigl( \mu, k_F \cos \theta - \frac{q}{2} \cos \phi, k_F \sin \theta - \frac{q}{2} \sin \phi \Bigr)$, 
$p_1^* = $$\Bigl( 3\mu, k_F \cos \theta' + \frac{Q}{2} \cos \Phi, k_F \sin \theta' + \frac{Q}{2} \sin \Phi \Bigr)$, 
$p_2^* = $$\Bigl( -\mu, -k_F \cos \theta' + \frac{Q}{2} \cos \Phi, -k_F \sin \theta' + \frac{Q}{2} \sin \Phi \Bigr)$, 
$p_3^* = $$\Bigl( \mu, -k_F \cos \theta + \frac{Q}{2} \cos \Phi, -k_F \sin \theta + \frac{Q}{2} \sin \Phi \Bigr)$, 
$p_4^* = $$\Bigl( \mu, k_F \cos \theta + \frac{Q}{2} \cos \Phi, k_F \sin \theta + \frac{Q}{2} \sin \Phi \Bigr)$.
The frequencies,
which are order of the floating energy scale $\mu$,
are chosen such that 
infrared divergences are cut off by $\mu$ in all particle-particle and particle-hole channels.
The spatial momenta lie on the Fermi surface 
when $\vec q$ and $\vec Q$ vanish.
$F_i$ are RG scheme dependent  corrections that are regular in the small $\mu$ limit.
The local counter terms needed to enforce the RG conditions is written as
\begin{eqnarray}
S_{CT}
&=& k_F 
\int \frac{ d \bar{\mathcal{V}}_2 }{(2\pi)^3}
~\sum_{\sigma =\pm} \psi^\dag_{\sigma;\theta}(\omega, \kappa ) (-i \delta_1 \omega + \delta_2 v_F \kappa  )\psi_{\sigma;\theta}(\omega, \kappa)  \nn
&+&
\frac{k_F^2}{4} 
 \int \frac{d\bar{\mathcal{V}}_4^{(0)}}{(2\pi)^9}  ~\sum_{\sigma_{1,2,3,4}=\pm}
(A^\lambda_0)^{ \theta-\vartheta^S_\theta, \sigma_1; ~ \theta'+\vartheta^S_{\theta'}, \sigma_2}_{ \theta+\vartheta^S_\theta, \sigma_4; ~ \theta'-\vartheta^S_{\theta'}, \sigma_3 } 
 (\lambda_0)^{ \theta-\vartheta^S_\theta, \sigma_1; ~ \theta'+\vartheta^S_{\theta'}, \sigma_2}_{ \theta+\vartheta^S_\theta, \sigma_4; ~ \theta'-\vartheta^S_{\theta'}, \sigma_3 }  
 \nonumber \\
&&\times
\psi^\dag_{\sigma_1;\theta-\vartheta^S_{\theta}}[\Omega+\omega/2,\kappa + k_F \vartheta_\theta^C ]
\psi^\dag_{\sigma_2; \theta'+\vartheta^S_{\theta'}}[\Omega'-\omega/2,\rho-k_F \vartheta_{\theta'}^C ]\nonumber\\
&&\times \psi_{\sigma_3; \theta'-\vartheta^S_{\theta'} }[\Omega'+\omega/2,\rho +k_F \vartheta^C_{\theta'}]
\psi_{\sigma_4; \theta+\vartheta^S_{\theta} }[\Omega-\omega/2,\kappa -k_F \vartheta^C_\theta] \nonumber \\
%%BCS
&+& 
\frac{k_F^2}{4} 
\int \frac{d\bar{\mathcal{V}}_4^{(1)}}{(2\pi)^9}  ~\sum_{\sigma_{1,2,3,4}=\pm} 
(A^\lambda_1)^{ \theta'-\varphi_{\theta'}^S, \sigma_1; ~ \theta' +\varphi^S_{\theta'}+\pi, \sigma_2}_{ \theta-\varphi^S_{\theta}, \sigma_4; ~ \theta+\varphi^S_{\theta}+\pi, \sigma_3 } 
(\lambda_1)^{ \theta'-\varphi_{\theta'}^S, \sigma_1; ~ \theta' +\varphi^S_{\theta'}+\pi, \sigma_2}_{ \theta-\varphi^S_{\theta}, \sigma_4; ~ \theta+\varphi^S_{\theta}+\pi, \sigma_3 } 
\nonumber \\
&&\times
\psi^\dag_{\sigma_1;\theta'-\varphi^S_{\theta'}}[
\Omega'+\omega/2,\rho + k_F \varphi^C_{\theta'}]
\psi^\dag_{\sigma_2; \theta'+\varphi^S_{\theta'}+\pi}[
-\Omega'+\omega/2,\rho-k_F \varphi^C_{\theta'} ]\nonumber\\
&&\times \psi_{\sigma_3; \theta+\varphi^S_{\theta}+\pi }[
-\Omega+\omega/2,\kappa -k_F \varphi^C_\theta]
\psi_{\sigma_4; \theta-\varphi^S_\theta}[
\Omega+\omega/2,\kappa +k_F \varphi^C_{\theta}].
\label{eq:counterterm}
\end{eqnarray}
Because the effective field theory is local, the RG 
 condition can be enforced with local counter terms.
The bare action becomes 
\begin{eqnarray}
S_B
&=& k_{F;B} \int 
\frac{d\bar{\mathcal{V}}_{2;B}}{(2\pi)^3}
 ~\sum_{\sigma =\pm} \psi^\dag_{B;\sigma;\theta}(\omega_B, \kappa ) (-i \omega_B +v_F \kappa  )\psi_{B;\sigma;\theta}(\omega_B, \kappa) \nonumber\\
&+&
\frac{k_{F;B}^2}{4} \int \frac{d\bar{\mathcal{V}}_{4;B}^{(0)}}{(2\pi)^9}  ~\sum_{\sigma_{1,2,3,4}=\pm}
(\lambda_{0;B})^{
\theta-\vartheta^S_{\theta;B}, \sigma_1;
~ \theta'+\vartheta^S_{\theta';B}, \sigma_2}_{
  \theta+\vartheta^S_{\theta;B}, \sigma_4;
~ \theta'-\vartheta^S_{\theta';B}, \sigma_3
 } \nonumber \\
&&\times
\psi^\dag_{B;\sigma_1;\theta-\vartheta^S_{\theta;B}}[\Omega_B+ \omega_B/2,\kappa + k_F \vartheta^C_{\theta;B} ]
\psi^\dag_{B;\sigma_2; \theta'+\vartheta^S_{\theta';B}}[\Omega_B'-\omega_B/2,\rho-k_F \vartheta^C_{\theta';B} ]\nonumber\\
&&\times \psi_{B;\sigma_3; \theta'-\vartheta^S_{\theta';B} }[\Omega_B'+\omega_B/2,\rho +k_F \vartheta^C_{\theta';B}]
\psi_{B;\sigma_4; \theta+\vartheta^S_{\theta;B} }[\Omega_B-\omega_B/2,\kappa -k_F \vartheta^C_{\theta;B}] \nonumber \\
%%BCS
&+& 
\frac{k_{F;B}^2}{4} \int \frac{d\bar{\mathcal{V}}^{(1)}_{4;B}}{(2\pi)^9} ~\sum_{\sigma_{1,2,3,4}=\pm}
(\lambda_{1;B})^{
\theta'-\varphi^S_{\theta';B}, \sigma_1;
~ \theta'+\varphi^S_{\theta';B}+\pi , \sigma_2}_{
 \theta-\varphi^S_{\theta;B}, \sigma_4;
~ \theta+\varphi^S_{\theta;B}+\pi, \sigma_3
 } \nonumber \\
&&\times
\psi^\dag_{B;\sigma_1;\theta'-\varphi^S_{\theta';B}}[
\Omega_B'+\omega_B/2,\rho + k_F \varphi^C_{\theta';B} ]
\psi^\dag_{B;\sigma_2; \theta'+\varphi^S_{\theta';B}+\pi}[
-\Omega_B'+\omega_B/2,\rho-k_F \varphi^C_{\theta';B} ]\nonumber\\
&&\times \psi_{B;\sigma_3; \theta+\varphi^S_{\theta;B}+\pi }[
-\Omega_B+\omega_B/2,\kappa -k_F \varphi^C_{\theta;B}]
\psi_{B;\sigma_4; \theta-\varphi^S_{\theta;B}}[
\Omega_B+\omega_B/2,\kappa +k_F \varphi^C_{\theta;B}].
\end{eqnarray} 
The bare variables\footnote{
Here we have
\begin{eqnarray}
d\bar{\mathcal{V}}_{2;B} &=& d\omega_B d\kappa d\theta    \nonumber\\
d\bar{\mathcal{V}}_{4;B}^{(0)} &=& d\omega_B d\Omega_B d\Omega'_B d\kappa d\rho d\phi dq q d\theta d\theta' \nonumber\\
d\bar{\mathcal{V}}_{4;B}^{(1)} &=& d\omega_B d\Omega_B d\Omega'_B d\kappa d\rho d\Phi dQ Q d\theta d\theta' \nonumber\\
\vartheta^{S}_{\theta,B}&=& \frac{q}{2k_{F;B} } \sin (\theta-\phi) ,\quad \varphi^{S}_{\theta,B}= \frac{Q}{2k_{F;B} }\sin(\theta-\Phi), \nn
\vartheta^{C}_{\theta,B}&=& \frac{q}{2k_{F;B} } \cos (\theta-\phi) ,\quad \varphi^{C}_{\theta,B}= \frac{Q}{2k_{F;B} }\cos(\theta-\Phi).
\end{eqnarray}
} are related to the 
renormalized variables through
the multiplicative renormalization factors,
$Z_i=1+\delta_i$, 
$(Z^\lambda_i)^{\theta_1, \theta_2}_{\theta_4,\theta_3} 
= 1+ 
(A^\lambda_i)^{\theta_1, \theta_2}_{\theta_4,\theta_3} 
$
for $i=1,2$,
\begin{eqnarray}
\omega_B = \frac{Z_1}{Z_2} \omega , \quad \psi_B = \sqrt{\frac{Z_2^2}{Z_1}}\psi , \quad 
(\lambda_{i;B})^{\theta_1, \theta_2}_{\theta_4,\theta_3} 
= 
\frac{
(Z^\lambda_i)^{\theta_1, \theta_2}_{\theta_4,\theta_3} 
\mu^{-1}
(\tilde \lambda_i)^{\theta_1, \theta_2}_{\theta_4,\theta_3} 
}{Z_1Z_2}, \quad
k_{B;F} = 
\mu \tilde k_{F}.  
\end{eqnarray}
Here, we use the scheme in which $v_F$ is fixed.
$(\tilde \lambda_i)^{\theta_1, \theta_2}_{\theta_4,\theta_3} =
\mu
( \lambda_i)^{\theta_1, \theta_2}_{\theta_4,\theta_3}$
and $\tilde k_F = \mu^{-1} k_F$ represent 
dimensionless objects
that are measured in the unit of the floating energy scale.
The beta functionals for the coupling functions are obtained by keeping $\lambda_{i;B}$ fixed with varying the floating energy.
This leads to the beta functionals,
\begin{eqnarray}
\frac{d  (\tilde \lambda_i)^{\theta_1, \theta_2}_{\theta_4,\theta_3}  }{dl}  =
\left[
-1 -3 (z-1) - 4 \eta_\psi+
\frac{ d \log 
(Z^\lambda_i)^{\theta_1, \theta_2}_{\theta_4,\theta_3} }{d \log \mu}
\right]
(\tilde \lambda_i)^{\theta_1, \theta_2}_{\theta_4,\theta_3}.
\end{eqnarray}
Here $l = \log (\Lambda /\mu)$ is the logarithmic length scale with
$\Lambda$ being a UV cutoff.
$z= \frac{ d \log Z_1/Z_2}{d \log \mu}$
is the dynamical critical exponent
and
$\eta_\psi= \frac{ d \log \sqrt{  Z_2^2/Z_1} }{d \log \mu}$
is the anomalous dimension of the fermion.
The dimensionless Fermi momentum obeys
$\frac{d  \tilde 
 k_F }{dl}  = \tilde k_F$.

At the one-loop order, only the following diagrams
contribute to the beta functionals,
\begin{eqnarray}
\Gamma^{(4)}
&=&
\begin{tikzpicture}[baseline={([yshift=-4pt]current bounding box.center)}]
\coordinate (v1) at (-15pt,-15pt);
\coordinate (v2) at (-15pt,15pt);
\coordinate (v3) at (-10pt,0pt);
\coordinate (v4) at (10pt,0pt);
\coordinate (v5) at (15pt,-15pt);
\coordinate (v6) at (15pt,15pt);
\draw[thick,postaction={mid arrow=red} ](v1)--  (v3);
\draw[thick,postaction={mid arrow=red} ](v3)--  (v2);
\draw[thick,postaction={mid arrow=red} ](v5)--  (v4);
\draw[thick,postaction={mid arrow=red} ](v4)--  (v6);
\draw[thick,postaction={mid arrow=red} ] (v3) to [bend left=30] (v4);
\draw[thick,postaction={mid arrow=red} ] (v4) to [bend left=30] (v3);
\node at (-22pt,-20pt) {\scriptsize $4$};
\node at (-22pt,20pt) {\scriptsize $1$};
\node at (22pt,-20pt) {\scriptsize $3$};
\node at (22pt,20pt) {\scriptsize $2$};
\node at (v1)[circle,fill,inner sep=1pt]{};
\node at (v2)[circle,fill,inner sep=1pt]{};
\node at (v5)[circle,fill,inner sep=1pt]{};
\node at (v6)[circle,fill,inner sep=1pt]{};
\end{tikzpicture} \quad 
+
\begin{tikzpicture}[baseline={([yshift=-4pt]current bounding box.center)}]
\coordinate (v1) at (-15pt,-15pt);
\coordinate (v2) at (-15pt,15pt);
\coordinate (v3) at (-10pt,0pt);
\coordinate (v4) at (10pt,0pt);
\coordinate (v5) at (15pt,-15pt);
\coordinate (v6) at (15pt,15pt);
\draw[thick,postaction={mid arrow=red} ](v1)--  (v3);
\draw[thick,postaction={mid arrow=red} ](v3)--  (v2);
\draw[thick,postaction={mid arrow=red} ](v5)--  (v4);
\draw[thick,postaction={mid arrow=red} ](v4)--  (v6);
\draw[thick,postaction={mid arrow=red} ] (v3) to [bend left=30] (v4);
\draw[thick,postaction={mid arrow=red} ] (v4) to [bend left=30] (v3);
\node at (-22pt,-20pt) {\scriptsize $3$};
\node at (-22pt,20pt) {\scriptsize $1$};
\node at (22pt,-20pt) {\scriptsize $4$};
\node at (22pt,20pt) {\scriptsize $2$};
\node at (v1)[circle,fill,inner sep=1pt]{};
\node at (v2)[circle,fill,inner sep=1pt]{};
\node at (v5)[circle,fill,inner sep=1pt]{};
\node at (v6)[circle,fill,inner sep=1pt]{};
\end{tikzpicture}
+\quad
\begin{tikzpicture}[baseline={([yshift=-4pt]current bounding box.center)}]
\coordinate (v1) at (-15pt,15pt);
\coordinate (v2) at (15pt,15pt);
\coordinate (v3) at (0pt,10pt);
\coordinate (v4) at (0pt,-10pt);
\coordinate (v5) at (-15pt,-15pt);
\coordinate (v6) at (15pt,-15pt);
\draw[thick,postaction={mid arrow=red} ](v3)--  (v1);
\draw[thick,postaction={mid arrow=red} ](v3)--  (v2);
\draw[thick,postaction={mid arrow=red} ](v5)--  (v4);
\draw[thick,postaction={mid arrow=red} ](v6)--  (v4);
\draw[thick,postaction={mid arrow=red} ] (v4) to [bend right=30] (v3);
\draw[thick,postaction={mid arrow=red} ] (v4) to [bend left=30] (v3);
\node at (-22pt,-20pt) {\scriptsize $3$};
\node at (-22pt,20pt) {\scriptsize $1$};
\node at (22pt,-20pt) {\scriptsize $4$};
\node at (22pt,20pt) {\scriptsize $2$};
\node at (v1)[circle,fill,inner sep=1pt]{};
\node at (v2)[circle,fill,inner sep=1pt]{};
\node at (v5)[circle,fill,inner sep=1pt]{};
\node at (v6)[circle,fill,inner sep=1pt]{};
\end{tikzpicture},
\label{eq:diagrams}
\end{eqnarray}
and $z=1$ and $\eta_\psi=0$.
The first diagram contributes to the near forward scatterings.
The third diagram contributes the pairing interaction.
In the following sections, 
we compute the beta functionals in the two channels
for general $\vec q$ and $\vec Q$.

\section{  Nearly forward scattering \label{sec:ZS}}

At the one-loop order, 
only the first diagram in Eq.~\eqref{eq:diagrams} 
contributes to the quantum correction of the near forward scattering processes.
The counter term reads 
\begin{eqnarray}
&& 
(A_0^\lambda)_{\theta+\vartheta_\theta^S,\sigma_4;\theta'-\vartheta_{\theta'}^S,\sigma_3}^{\theta-\vartheta_\theta^S,\sigma_1;\theta'+\vartheta_{\theta'}^S,\sigma_2} 
(\lambda_0)_{\theta+\vartheta_\theta^S,\sigma_4;\theta'-\vartheta_{\theta'}^S,\sigma_3}^{\theta-\vartheta_\theta^S,\sigma_1;\theta'+\vartheta_{\theta'}^S,\sigma_2} 
= \nn
&& 
-\frac{1}{2 \mu} 
\frac{k_F}{\mu}
\int \frac{d\theta''  d\kappa'' }{(2\pi)^2} \frac{d\Omega''}{2\pi}\sum_{\sigma',\sigma''}~
(\tilde{\lambda}_0)_{\theta+\vartheta_\theta^S,\sigma_4;\theta''-\vartheta_{\theta''}^S,\sigma'}^{\theta-\vartheta_\theta^S,\sigma_1;\theta''+\vartheta_{\theta''}^S,\sigma''} 
(\tilde{\lambda}_0)_{\theta''+\vartheta_{\theta''}^S,\sigma'';\theta'-\vartheta_{\theta'}^S,\sigma_3}^{\theta''-\vartheta_{\theta''}^S,\sigma';\theta'+\vartheta_{\theta'}^S,\sigma_2}
\nonumber\\
&& 
~~~~~ \times~ 
Re \left[
\frac{1}{-i\Omega''+v_F \left[\kappa''
-\frac{q}{2}\cos(\theta''-\phi)
\right]}\frac{1}{-i(\mu+\Omega'')+v_F \left[\kappa''
+ \frac{q}{2}\cos(\theta''-\phi)
\right]}
\right].
\end{eqnarray}
The integrations over $\kappa''$ and $\Omega''$ 
can be readily done.
Apart from the overall factor of $1/\mu$
determined from the dimension of the quartic coupling,
the counter term is proportional to $k_F/\mu$.
This reflects the fact that
the phase space of the virtual particle-hole pairs
is proportional to $k_F$.
The phase space 
measured in the unit $\mu$
increases with decreasing $\mu$.
This extensive phase space is 
what promotes the quartic coupling 
to the marginal coupling
although it has scaling dimension $-1$\cite{BORGES2023169221}.
One can incorporate the phase space 
to define a new dimensionless coupling function as
$[F_{\theta,\theta'}(q,\phi)]^{\sigma_1;\sigma_2}_{\sigma_4;\sigma_3}= \frac{k_F}{\mu}
(\tilde{\lambda}_0)_{\theta+\vartheta_\theta^S,\sigma_4;\theta'-\vartheta_{\theta'}^S,\sigma_3}^{\theta-\vartheta_\theta^S,\sigma_1;\theta'+\vartheta_{\theta'}^S,\sigma_2}$. 
The $\beta$ functional for $F$  becomes
independent of $k_F$,
\begin{eqnarray}
\frac{
dF^{r}_{\theta_1,\theta_2}(q,\phi)
}{dl} = -\frac{1}{4\pi^2v_F  }  \int_0^{2\pi} d\theta ~F^{r}_{ \theta_1,\theta}(q,\phi)~F^{r}_{\theta,\theta_2}(q,\phi) ~
\frac{\mu^2 v_F^2q^2\cos^2(\theta-\phi)}{[\mu^2+v_F^2q^2\cos^2 (\theta-\phi)]^2}.
\label{eq:beta_ZS}
\end{eqnarray}
Here, 
$r=s$ or $a$.
$F^{s}_{\theta_1,\theta_2}(q,\phi)$
and
$F^{a}_{\theta_1,\theta_2}(q,\phi)$
represent
the projections of 
$[F_{\theta,\theta'}(q,\phi)]^{\sigma_1;\sigma_2}_{\sigma_4;\sigma_3}$
to the singlet and adjoint representations 
of a particle-hole pair for the $SU(2)$ group, respectively,
\begin{eqnarray}
[F_{\theta_1,\theta_2}(q,\phi)]_{\sigma_4;\sigma_3}^{\sigma_1;\sigma_2} =\mathfrak{S}_{\sigma_4,\sigma_3}^{\sigma_1,\sigma_2} F^s_{\theta_1,\theta_2}(q,\phi)+\mathfrak{A}_{\sigma_4,\sigma_3}^{\sigma_1,\sigma_2}
F^a_{\theta_1,\theta_2}(q,\phi),
\end{eqnarray}
where 
$\mathfrak{S}_{\sigma_4,\sigma_3}^{\sigma_1,\sigma_2} =\frac{1}{2}\delta^{\sigma_1}_{\sigma_4}\delta^{\sigma_2}_{\sigma_3}$ 
and
$\mathfrak{A}_{\sigma_4,\sigma_3}^{\sigma_1,\sigma_2} =\delta^{\sigma_1}_{\sigma_3}\delta^{\sigma_2}_{\sigma_4}-\frac{1}{2}\delta^{\sigma_1}_{\sigma_4}\delta^{\sigma_2}_{\sigma_3}$.
From now on, we omit the superscript $r$ because 
the following analysis holds for both channels.

The beta functional in \eq{eq:beta_ZS} vanishes at $q=0$
for any non-zero $\mu$.
This is consistent with the fact that the
strict forward scattering amplitude is exactly marginal.
However, for any non-zero $\mu$,
the beta functional is non-trivial 
 for $q \neq 0$.
This has interesting consequences.
First, when the theory flows to a fixed point in the low-energy limit,
the full coupling function of the local effective field theory
exhibits a scale invariance when the momentum transfer
is comparable with energy scale.
Second, the non-trivial renormalization group flow 
of the non-forward scattering amplitude
can create instabilities 
if the bare coupling is sufficiently negative.
In the following, 
we discuss these consequences in detail.

For a fixed momentum transfer,
we can view the coupling $F$ 
as a matrix of two angles
 that play the role
of continuous indices.
Here, the product of two matrices is given by $({\bf A}\cdot {\bf B})_{\theta_1,\theta_2}
=\int \frac{d\theta}{2\pi} {\bf A}_{\theta_1 \theta}
{\bf B}_{\theta \theta_1}
$
and
${\bf M}^{-1}$ denotes the inverse of matrix ${\bf M}$.
Then, Eq.~\eqref{eq:beta_ZS} can be cast into 
$\frac{d{\bf F}}{dl} = - {\bf F} \cdot {\bf D} \cdot {\bf F}$,
where ${\bf D}_{\theta, \theta'}
= -\frac{1}{v_F  }  
\frac{\mu^2 v_F^2q^2\cos^2(\theta-\phi)}{[\mu^2+v_F^2q^2\cos^2 (\theta-\phi)]^2}
\delta(\theta-\theta')
$.
Multiplying ${\bf F}^{-1}$ on both sides of 
Eq.~\eqref{eq:beta_ZS},
we obtain
\begin{eqnarray}
\frac{d[{\bf F}(q,\phi)]^{-1}_{\theta,\theta'}}{dl} &=& \frac{1}{v_F}\delta(\theta-\theta')  \frac{\mu^2 v_F^2q^2\cos^2(\theta-\phi)}{[\mu^2+v_F^2q^2\cos^2 (\theta-\phi)]^2}.
\label{eq:beta_inverse_ZS}
\end{eqnarray}
At scale $\mu$, the beta functional is largest
for $q \cos(\theta-\phi) \sim \mu$.
For $\theta-\phi \sim \pi/2$, 
the phase space of $q$ is largest
because  virtual particle-hole pairs with momentum $\vec q$ 
cost the least energy in the region of the Fermi surface 
where $\vec q$ is tangential to the Fermi surface.
The solution of Eq.~\eqref{eq:beta_inverse_ZS} is given by
\begin{eqnarray}
[{\bf F}(q,\phi;\mu)]^{-1}_{\theta,\theta'} &=& \frac{\delta(\theta-\theta')}{2 v_F } \left[\frac{v_F^2q^2\cos^2(\theta-\phi)}{\mu^2+v_F^2q^2\cos^2(\theta-\phi)}-\frac{v_F^2q^2\cos^2(\theta-\phi)}{\Lambda^2+v_F^2q^2\cos^2(\theta-\phi)} \right]+[{\bf F}(q,\phi;\Lambda)]^{-1}_{\theta,\theta'}, \nn
\label{eq:Fqphimu}
\end{eqnarray}
where ${\bf F}(q,\phi;\Lambda)$ is the coupling function at the UV cutoff, $\Lambda$. 
One can combine the last two terms to define the coupling function at $\mu = \infty$ to simplify the solution as
\begin{eqnarray}
[{\bf F}(q,\phi;\mu)]^{-1}_{\theta,\theta'} &=& \frac{1}{2 v_F }\delta(\theta-\theta')  \frac{v_F^2q^2\cos^2(\theta-\phi)}{\mu^2+v_F^2q^2\cos^2(\theta-\phi)} +[{\bf F}(q,\phi;\infty)]^{-1}_{\theta,\theta'}.
\label{eq:solution_F_theta}
\end{eqnarray}

\subsection{Scale invariance of Fermi liquid fixed points
\label{sec:FL}}

Let us first consider the case in which
no eigenvalue of ${\bf F}$ diverges
at any $\mu$.
In this case, the theory is expected to flow
to a Fermi liquid fixed point in the low-energy limit
if the interaction is repulsive in the pairing channel.
It is noted that 
the $\mu \rightarrow 0$ limit 
and the $q \rightarrow 0$ limit do not commute.
In the strict forward scattering limit ($q=0$),
the coupling function does not depend on $\mu$ at all.
On the other hand, if one takes the $\mu \rightarrow 0$ limit for a 
fixed $q \neq 0$, the coupling function saturates to
$[{\bf F}(q,\phi;0)]^{-1}_{\theta,\theta'} = \frac{1}{8\pi^2 v_F }\delta(\theta-\theta') +[{\bf F}(q,\phi;\infty)]^{-1}_{\theta,\theta'}$
in the low-energy limit.
The non-trivial crossover between these two limits 
can be captured by the scale invariant coupling function 
defined by
$[\tilde{\bf F}(\tilde{q},\phi;\mu)]_{\theta,\theta'}
\equiv
[{\bf F}( \mu \tilde q,\phi;\mu)]_{\theta,\theta'}$,
where $\tilde q = q/\mu$ corresponds to the dimensionless
momentum transfer measured in the unit of $\mu$.
This allows us to probe the kinematic region 
with small but non-zero momentum transfers.
By taking the $\mu \rightarrow 0$ limit with fixed $\tilde q$,
the coupling function becomes
\begin{eqnarray}
\lim_{\mu \rightarrow 0}
[\tilde{\bf F}(\tilde{q},\phi;\mu)]^{-1}_{\theta,\theta'} &=& \frac{\delta(\theta-\theta')}{2v_F }\frac{v_F^2 \tilde{q}^2\cos^2 (\theta-\phi)}{1+v_F^2 \tilde{q}^2\cos^2 (\theta-\phi)}+[{\bf F}(0,0;\infty)]^{-1}_{\theta,\theta'}.
\label{eq:beta_solution_rescale}
\end{eqnarray}
Here, we use the fact that 
$[{\bf F}(q,\phi;\infty)]_{\theta,\theta'}$ 
is an analytic function of $\vec q$.
In the $\mu \rightarrow 0$ limit,
only $[{\bf F}(q=0,\phi=0;\infty)]_{\theta,\theta'}$ 
enters in the expression for the fixed point coupling function.
\eq{eq:beta_solution_rescale}
corresponds to the fixed point of 
a beta functional,
$\left. \frac{d [\tilde{\bf F}(\tilde{q},\phi)]_{\theta,\theta'}}{d \ell } \right|_{\tilde q, \theta, \theta'}
 =  
 \left.
 \frac{d [{\bf F}(q,\phi)]_{\theta,\theta'}}{d \ell}
\right|_{q,\theta,\theta'} 
 - \tilde{q} \frac{\partial [\tilde{\bf F}(\tilde{q},\phi)]_{\theta,\theta'}}{\partial \tilde{q}}$,
where the second term in the new beta functional corresponds to 
a momentum dilatation that `magnifies' the region with small momentum transfers as the low-energy limit is taken.
It is noted that the scale invariant coupling function obeys the $z=1$ scaling.
The momentum transfer and energy scale in the same way because the energy of a particle-hole pair 
with momentum $\vec q= q (\cos \phi, \sin \phi)$ 
created near the Fermi surface at angle $\theta$
scales linearly in $q$ as far as 
$\theta-\phi \neq \pi/2$.
\eq{eq:beta_solution_rescale} is the central result of our paper.
The local coupling function 
captures the full extent of 
the scale invariant Fermi liquid fixed point away from the strict forward scattering limit.
As expected, the only UV information that is kept in the fixed point coupling is the forward scattering amplitude at $\vec q=0$.

In general, 
it is not easy to invert 
\eq{eq:beta_solution_rescale} 
to write down 
$[\tilde{\bf F}(\tilde{q},\phi;\mu)]_{\theta,\theta'}$ 
in a closed form\footnote{
If the UV coupling is weak,
we can write down the fixed point coupling function in powers of the UV coupling function as
$[\tilde{\bf F}(\tilde{q},\phi;\mu)]_{\theta,\theta'} = [{\bf F}(0,\phi;\infty)]_{\theta,\theta'}
-\frac{1}{8\pi^2 v_F }\int_0^{2\pi}d\Theta [{\bf F}(0,\phi;\infty)]_{\theta,\Theta} \frac{v_F^2 \tilde{q}^2\cos^2 (\Theta-\phi)}{1+v_F^2 \tilde{q}^2\cos^2 (\Theta-\phi)} [{\bf F}(0,\phi;\infty)]_{\Theta,\theta'} +\dots$.
}.
If the UV coupling function is non-zero only in one angular momentum channel,
the coupling function in the IR limit can be easily obtained.
To see this, we start by writing \eq{eq:Fqphimu}
in the space of angular momentum, 
\begin{eqnarray}
[{\bf F}(q,\phi;\mu)]^{-1}_{\ell,\ell'} &=& 
\int  \frac{d\theta}{8\pi^2v_F }\left[\frac{v_F^2 {q}^2\cos^2 (\theta-\phi)}{\mu^2+v_F^2 {q}^2\cos^2 (\theta-\phi)}-\frac{v_F^2 {q}^2\cos^2 (\theta-\phi)}{\Lambda^2+v_F^2 {q}^2\cos^2 (\theta-\phi)}\right]e^{i(\ell-\ell') \theta }\nonumber\\
&+&[{\bf F}(q,\phi;\Lambda)]^{-1}_{\ell,\ell'},
\label{eq:solution_F_angular}
\end{eqnarray}
where $\ell$ and $\ell'$ are conjugate momenta associated with $\theta$ and $\theta'$, respectively,
and
$
{\bf M}_{\ell, \ell'}= 
\int  \frac{d\theta d \theta'}{(2 \pi)^2}
{\bf M}_{\theta, \theta'} 
e^{i(\ell \theta -\ell' \theta')}
$.
In the angular momentum basis,
the fixed point coupling function can be written as
\begin{eqnarray}
    {\bf F}(q,\phi;\mu) 
    &=& [{\bf F}(q,\phi;\Lambda)]\Big( [
    \boldsymbol{\mathcal{ D}}'(q,\phi;\mu,\Lambda)
    {\bf F}(q,\phi;\Lambda)] + I \Big)^{-1},
\end{eqnarray}
where 
$
\boldsymbol{\mathcal{ D}}'_{\ell, \ell'}(q,\phi;\mu,\Lambda)
= e^{i (\ell-\ell') \phi}
{\mathcal{ D}}_{\ell, \ell'}(q;\mu,\Lambda)$
with
${\mathcal{ D}}_{\ell, \ell'}(q;\mu,\Lambda)=
\left[ D_{\ell-\ell'}(q;\mu)-D_{\ell-\ell'}(q;\Lambda) \right]$ 
and
\begin{eqnarray}
D_{\ell-\ell'}(q;\mu) 
&=& \frac{1}{
8\pi^2 
v_F } \int d\theta ~e^{i\theta (\ell-\ell')} \frac{ ~v_F^2 q^2 \cos^2(\theta)}{\mu^2+v_F^2 q^2 \cos^2(\theta)}.
\end{eqnarray}
$D_{\ell}(q,\mu)$ vanishes for odd $\ell$.
For $\ell=0,2,4$,
it takes the form of
\begin{eqnarray}
D_{0}(q;\mu) &=& \frac{1}{
4\pi 
v_F } \left[1-\frac{1}{\sqrt{1+\frac{v_F^2q^2}{\mu^2}}}\right], \nonumber\\
D_{ 2}(q;\mu) &=& \frac{1}{
4\pi 
v_F } \frac{\mu^2}{v_F^2q^2}\left[\sqrt{1+\frac{v_F^2q^2}{\mu^2}}+\frac{1}{\sqrt{1+\frac{v_F^2q^2}{\mu^2}}} -2\right], \nonumber\\
D_{4}(q;\mu) &=& \frac{1}{
4\pi 
v_F } \frac{\mu}{v_Fq}\left[\frac{4\mu}{v_Fq}\Big(1+\frac{2\mu^2}{v_F^2q^2}-\frac{2\mu}{v_Fq}\sqrt{\frac{\mu^2}{v_F^2q^2}+1}\Big)-\frac{1}{\sqrt{\frac{\mu^2}{v_F^2q^2}+1}}\right].
\end{eqnarray}

\subsubsection{s-wave}

\begin{figure}[h]
    \centering
    \includegraphics[width=.9\textwidth]{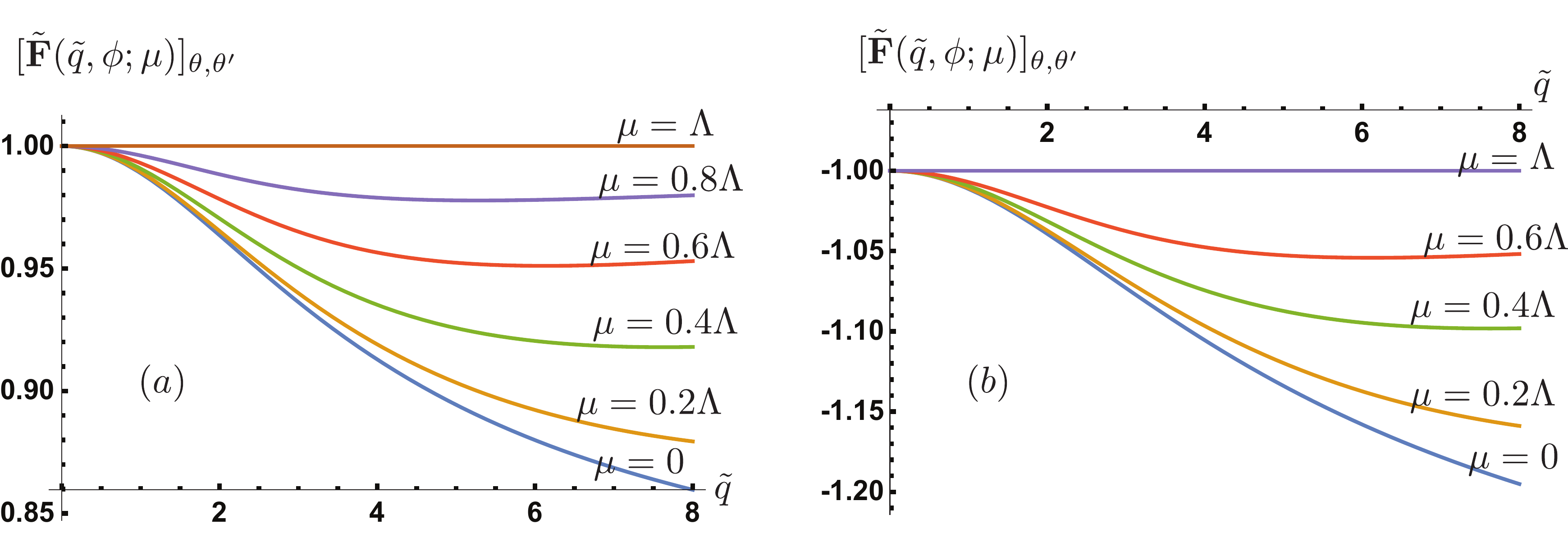}
    \caption{
RG flow of 
    $[\tilde{\bf F}(\tilde{q},\phi;\mu)]_{\theta,\theta'}$ 
for a momentum-independent 
UV coupling function
$[{\bf F}(q,\phi;\Lambda)]_{\theta,\theta'} =\alpha$
with
(a) $\alpha=1$ and 
(b) $\alpha=-1$.
For both plots, we choose $v_F=0.3$ and $\Lambda=10$.
For the angle-independent UV coupling function,
    $[\tilde{\bf F}(\tilde{q},\phi;\mu)]_{\theta,\theta'}$ 
    remains independent of
 $\phi$, $\theta$ and $\theta'$
 at all scales.
 The coupling function in the  $\mu \rightarrow 0$ limit represents the fixed point profile.
    }
    \label{fig:zs_RG_flow_FL}
\end{figure}

Let us first consider the case 
where the forward scattering amplitude 
of the UV coupling is 
independent of angles,
$[{\bf F}(q,\phi;\Lambda)]_{\theta,\theta'} =\alpha$,
where $\alpha$ denotes the strength of the coupling in the s-wave channel.
In the basis of angular momentum, 
the UV coupling is written as
$[{\bf F}(q,\phi;\Lambda)]_{\ell,\ell'}=\alpha \delta_{\ell,0}\delta_{\ell',0}$.
Using
$[ \boldsymbol{\mathcal{D}}'(q,\phi;\mu,\Lambda) {\bf F}(q,\phi;\Lambda) ]_{\ell,\ell'} 
=\alpha  \mathcal{D}'_{\ell,0}(q,\phi;\mu,\Lambda)  \delta_{\ell',0}$,
we can write
$ \boldsymbol{\mathcal{ D}}'(q,\phi;\mu,\Lambda) {\bf F}(q,\phi;\Lambda)+I$ as
\begin{eqnarray}
\boldsymbol{\mathcal{D}}'(q,\phi;\mu,\Lambda) {\bf F}(q,\phi;\Lambda)+I &=&  \left[\begin{array}{ccccc}
   1  & 0 & \alpha  \mathcal{D}'_{-2}(q,\phi;\mu,\Lambda) & 0 & 0 \\
   0 & 1 & 0 & 0 & 0 \\
   0 & 0 & 1+\alpha\mathcal{D}'_{0}(q,\phi;\mu,\Lambda) & 0 & 0 \\
   0 & 0 & 0 & 1 & 0 \\
   0  & 0 & \alpha  \mathcal{D}'_{2}(q,\phi;\mu,\Lambda) & 0 & 1 \\
\end{array}\right], 
\end{eqnarray}
where the matrix elements 
 are explicitly shown 
 only in the $5 \times 5$ block of $-2 \leq \ell,\ell' \leq 2$.
This leads to the isotropic IR quartic coupling function, 
$    
{\bf F}(q,\phi;\mu)
    ={\bf F}(q,\phi;\Lambda)\left[\boldsymbol{\mathcal{ D}}'(q,\phi;\mu,\Lambda) {\bf F}(q,\phi;\Lambda)+I\right]^{-1} =
\delta_{\ell,0}\delta_{\ell',0}  F(q,\phi;\mu)
$,
where
$F(q,\phi;\mu) =
\frac{\alpha}{\alpha \mathcal{D}_{0,0}(q;\mu,\Lambda)+1}= 
\alpha \left\{
\frac{\alpha}{
4\pi 
v_F}\left[\frac{1}{\sqrt{1+\frac{v_F^2q^2}{\Lambda^2}}}-\frac{1}{\sqrt{1+\frac{v_F^2q^2}{\mu^2}}} \right]  + 1 
\right\}^{-1}$. 
In the space of angles, we readily obtain
\bqa
[{\bf F}(q,\phi;\mu)]_{\theta,\theta'} =\frac{\alpha}{
\frac{\alpha}{
4\pi 
v_F}\left[\frac{1}{\sqrt{1+\frac{v_F^2q^2}{\Lambda^2}}}-\frac{1}{\sqrt{1+\frac{v_F^2q^2}{\mu^2}}} \right]  + 1 }. 
\label{eq:FwithLambda}
\eqa
In the large 
 $\Lambda$ limit,
 it takes a simpler form,
% letting $\alpha=F(q,\phi;\infty)$, we have
\begin{eqnarray}
   F(q,\phi;\mu) = 
\alpha \left\{
\frac{\alpha}{
4\pi 
v_F}\left[1-\frac{1}{\sqrt{1+\frac{v_F^2q^2}{\mu^2}}} \right]  + 1 
\right\}^{-1}.
\label{eq:F_s_wave}
\end{eqnarray}
Fig.~\ref{fig:zs_RG_flow_FL} shows the evolution of coupling functions.
In the low-energy limit,
the coupling function converges to
the fixed point profile,
\begin{eqnarray}
[\tilde{\bf F}(\tilde q,\phi;0)]_{\theta, \theta'} =  
\frac{\alpha}{\frac{\alpha}{
4\pi 
v_F}\left[
1
-\frac{1}{\sqrt{1+v_F^2\tilde{q}^2}} \right]  + 1}.
\label{eq:solution_F_Lambda_s-wave_rescaling}
\end{eqnarray}

\subsubsection{d-wave}

\begin{figure}[h]
    \centering
    \includegraphics[width=.9\textwidth]{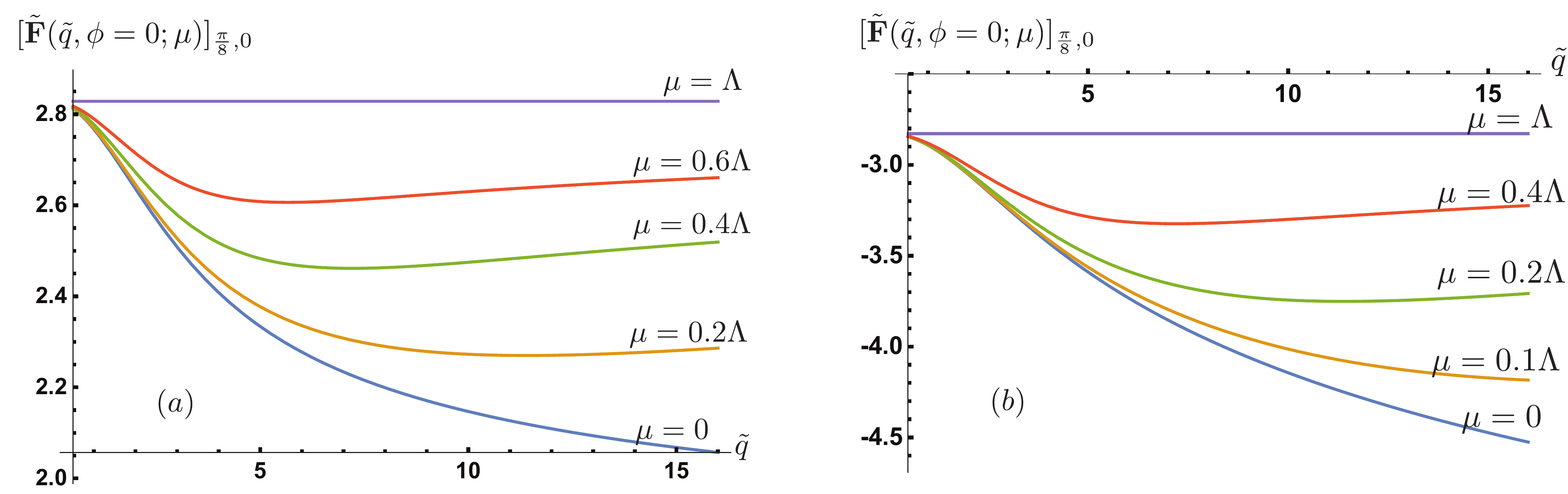}
    \caption{
RG flow of 
    $[\tilde{\bf F}(\tilde{q},\phi=0;\mu)]_{\frac{\pi}{8},0}$ 
for a d-wave UV coupling 
 function  
 $[{\bf F}(q,\phi;\Lambda)]_{\theta,\theta'} =2\alpha \cos [2(\theta-\theta')]$
for (a) $\alpha=2$,
and (b) $\alpha=-2$.
We use $v_F=0.3$ and $\Lambda=10$.
    }
    \label{fig:evolve_f_dwave_FL}
\end{figure}

\begin{figure}[h]
    \centering
    \includegraphics[width=.9\textwidth]{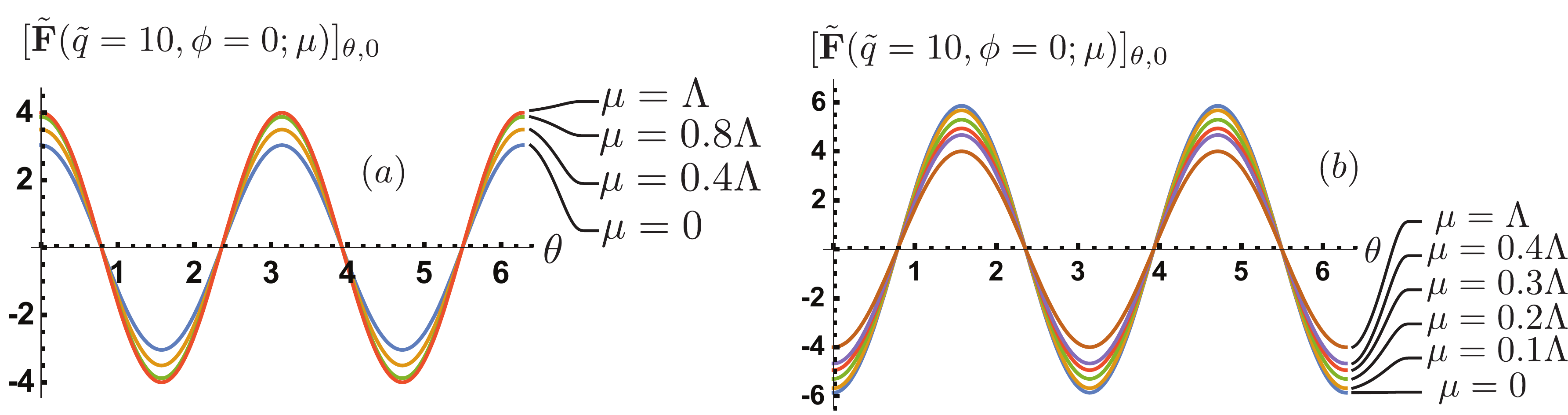}
    \caption{
    $[\tilde{\bf F}(\tilde{q}=10,\phi=0;\mu)]_{\theta,0}$ plotted as a function of $\theta$ 
    at different values of $\mu$ for (a) $\alpha=2$ 
    and (b) $\alpha=-2$.
   The same UV coupling function and parameters are used as in
    Fig. \ref{fig:evolve_f_dwave_FL}.
    }
    \label{fig:evolve_f_dwave_theta_FL}
\end{figure}

As the next example,
let us consider the case where
the UV coupling has only the d-wave component,
$[{\bf F}(q,\phi;\Lambda)]_{\theta,\theta'} =2\alpha \cos [2(\theta-\theta')]$.
In the angular momentum basis, 
the UV coupling function can be written as
$[{\bf F}(q,\phi;\Lambda)]_{\ell,\ell'} 
= \alpha(\delta_{\ell,-2}\delta_{\ell',-2}+\delta_{\ell,2}\delta_{\ell',2})$.
From
\begin{eqnarray}
\sum_{l''}  \boldsymbol{ \mathcal{D}}'_{\ell,\ell''}(q,\phi;\mu,\Lambda)[{\bf F}(q,\phi;\Lambda)]_{\ell'',\ell'} =
\alpha\left[ \boldsymbol{ \mathcal{D}}'_{\ell,-2} (q,\phi;\mu,\Lambda)\delta_{\ell', -2}  
+  
\boldsymbol{ \mathcal{D}}'_{\ell,2}(q,\phi;\mu,\Lambda) \delta_{\ell', 2}\right],
\end{eqnarray}
we obtain  the $5 \times 5$ block of
$ \boldsymbol{\mathcal{D}}'(q,\phi;\mu,\Lambda){\bf F}(q,\phi;\Lambda) +I $ 
for $-2 \leq \ell,\ell' \leq 2$ as
\begin{eqnarray}
  \boldsymbol{\mathcal{D}}'(q,\phi;\mu,\Lambda){\bf F}(q,\phi;\Lambda) +I &=&  \left[\begin{array}{ccccc}
   1+\alpha\mathcal{D}'_{0}(q,\phi;\mu,\Lambda)  & 0 & 0 & 0 & \alpha  \mathcal{D}'_{-4}(q,\phi;\mu,\Lambda) \\
   0 & 1 & 0 & 0 & 0 \\
   \alpha  \mathcal{D}'_{2}(q,\phi;\mu,\Lambda) & 0 & 1 & 0 & \alpha \mathcal{D}'_{-2}(q,\phi;\mu,\Lambda) \\
   0 & 0 & 0 & 1 & 0 \\
   \alpha  \mathcal{D}'_{4}(q,\phi;\mu,\Lambda)  & 0 & 0 & 0 & 1+\alpha \mathcal{D}'_{0}(q,\phi;\mu,\Lambda) \\
\end{array}\right]. 
\end{eqnarray}
The non-zero components of the IR coupling function is obtained to be
\begin{eqnarray}
  && [{\bf F} (q,\phi;\mu)]_{\ell,\ell'}  =  \nn
  && \alpha
   \frac{\left[1+\alpha  \mathcal{D}_{0}(q;\mu,\Lambda)\right](\delta_{\ell,2}\delta_{\ell',2}+\delta_{\ell,-2}\delta_{\ell,-2})-\alpha \mathcal{D}_{4}(q;\mu,\Lambda)(e^{4i\phi}\delta_{\ell,2}\delta_{\ell',-2}+e^{-4i\phi}\delta_{\ell,-2}\delta_{\ell',2})}{1+ 2\alpha \mathcal{D}_{0}(q;\mu,\Lambda)+\alpha^2\left[ \mathcal{D}_{0}^2 (q;\mu,\Lambda)- \mathcal{D}_{4}^2(q;\mu,\Lambda)\right]}.
\end{eqnarray}
This can be readily transformed back to the angle basis as
\begin{eqnarray}
[{\bf F}(q,\phi;\mu) ]_{\theta,\theta'}
&=& 2\alpha  \frac{\left[1+\alpha \mathcal{D}_{0}(q;\mu,\Lambda)\right]\cos [2(\theta-\theta')]-\alpha \mathcal{D}_{4}(q;\mu,\Lambda)\cos [2(\theta+\theta')-4\phi]}{1+ 2\alpha \mathcal{D}_{0}(q;\mu,\Lambda)+\alpha^2\left[ \mathcal{D}_{0}^2(q;\mu,\Lambda) - \mathcal{D}_{4}^2(q;\mu,\Lambda)\right]}.
\end{eqnarray}
In the $\mu \rightarrow 0$ limit, 
the coupling function takes the universal form given by
\begin{eqnarray}
[\tilde{\bf F}(\tilde{q},\phi;0)]_{\theta,\theta'} =2\alpha \frac{ [1+\alpha  \tilde{D}_0(\tilde{q})] \cos [2(\theta-\theta')]-\alpha  \tilde{D}_4(\tilde{q}) \cos [2(\theta+\theta')-4\phi]}{1+2\alpha  \tilde{D}_0(\tilde{q})+\alpha^2 [ \tilde{D}_0^2(\tilde{q})- \tilde{D}_4^2(\tilde{q})] },
\label{eq:f_d_wave_IR}
\end{eqnarray}
where  
$\tilde{q}=q/\mu$
and
\begin{eqnarray}
\tilde{D}_0 (x)&=& \frac{1}{
4\pi v_F} \left[1-\frac{1}{\sqrt{1+v_F^2x^2}}\right], \nonumber\\
\tilde{D}_4 (x) &=& \frac{1}{
4\pi v_F} \left[-\frac{1}{\sqrt{1+v_F^2x^2}} +\frac{4}{v_F^2x^2}\Big(1+\frac{2}{v_F^2x^2}-\frac{2}{v_F x}\sqrt{1+\frac{1}{v_F^2x^2}}\Big)\right].
\end{eqnarray}
The evolution of the coupling function is shown in Fig.~\ref{fig:evolve_f_dwave_FL}. 
In Fig.~\ref{fig:evolve_f_dwave_theta_FL}, 
the coupling function is shown 
as a function of angle at different energy scales.

\subsection{
Instabilities of Fermi liquids 
in the particle-hole channel
\label{sec:instability}}

So far, we have considered the cases
where the theory flows to the Fermi liquid fixed point
in the low energy limit.
If the UV coupling is sufficiently attractive 
in one or more angular momentum channel,
some eigenvalues of the coupling function 
can diverges at low energies,
signifying potential instabilities.
However, this instability in the particle-hole channel requires a finite strength of coupling.
While the perturbative analysis is not 
 expected to be quantitatively valid,
 the main point of this analysis is to highlight the importance 
 of quantum corrections to non-forward scatterings
 for particle-hole instabilities.

\subsubsection{the s-wave channel}

\begin{figure}[h]
    \centering
    \includegraphics[width=.9\textwidth]{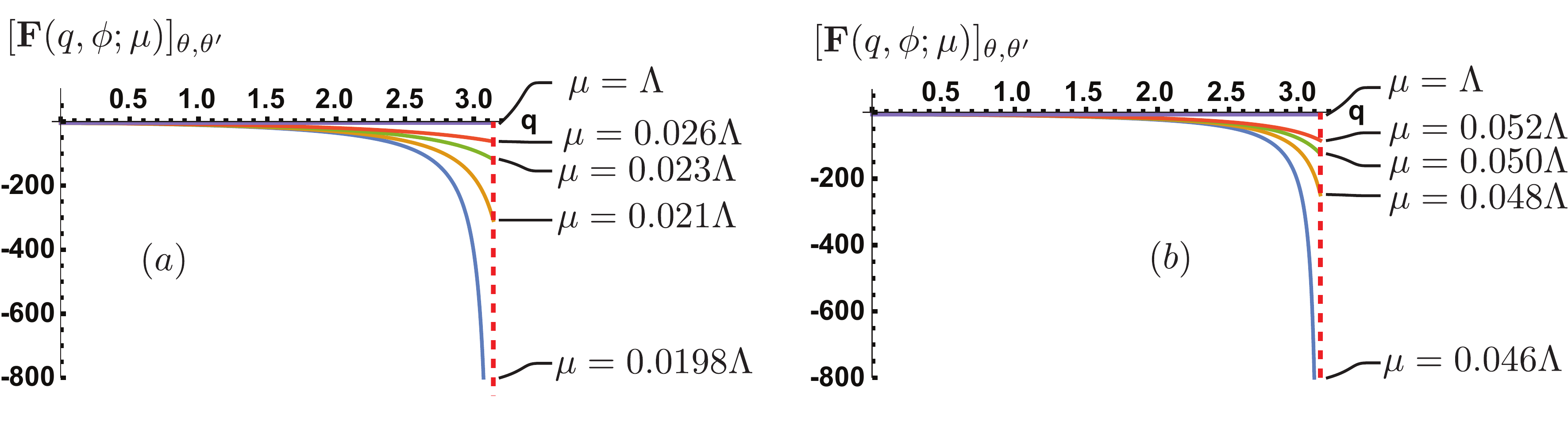}
    \caption{
    RG flow of $[{\bf F}({q},\phi;\mu)]_{\theta,\theta'}$ 
   for an attractive UV coupling function 
    $[{\bf F}(q,\phi;\Lambda)]_{\theta,\theta'} =\alpha$
    with 
    (a) $\alpha=-(4\pi v_F+1)$ 
    and
    (b) $\alpha=-(4\pi v_F+3)$. 
   The coupling function diverges at the momentum cutoff,
   which is chosen to be 
   $q_c=\pi$,
   at $\mu_c=0.0198\Lambda$ 
   and $\mu_c=0.046\Lambda$, respectively.
    We use $\Lambda=10$ and $v_F=0.3$.
    }
    \label{fig:zs_RG_flow_Instability}
\end{figure}

\begin{figure}[h]
    \centering
    \includegraphics[width=.47\textwidth]{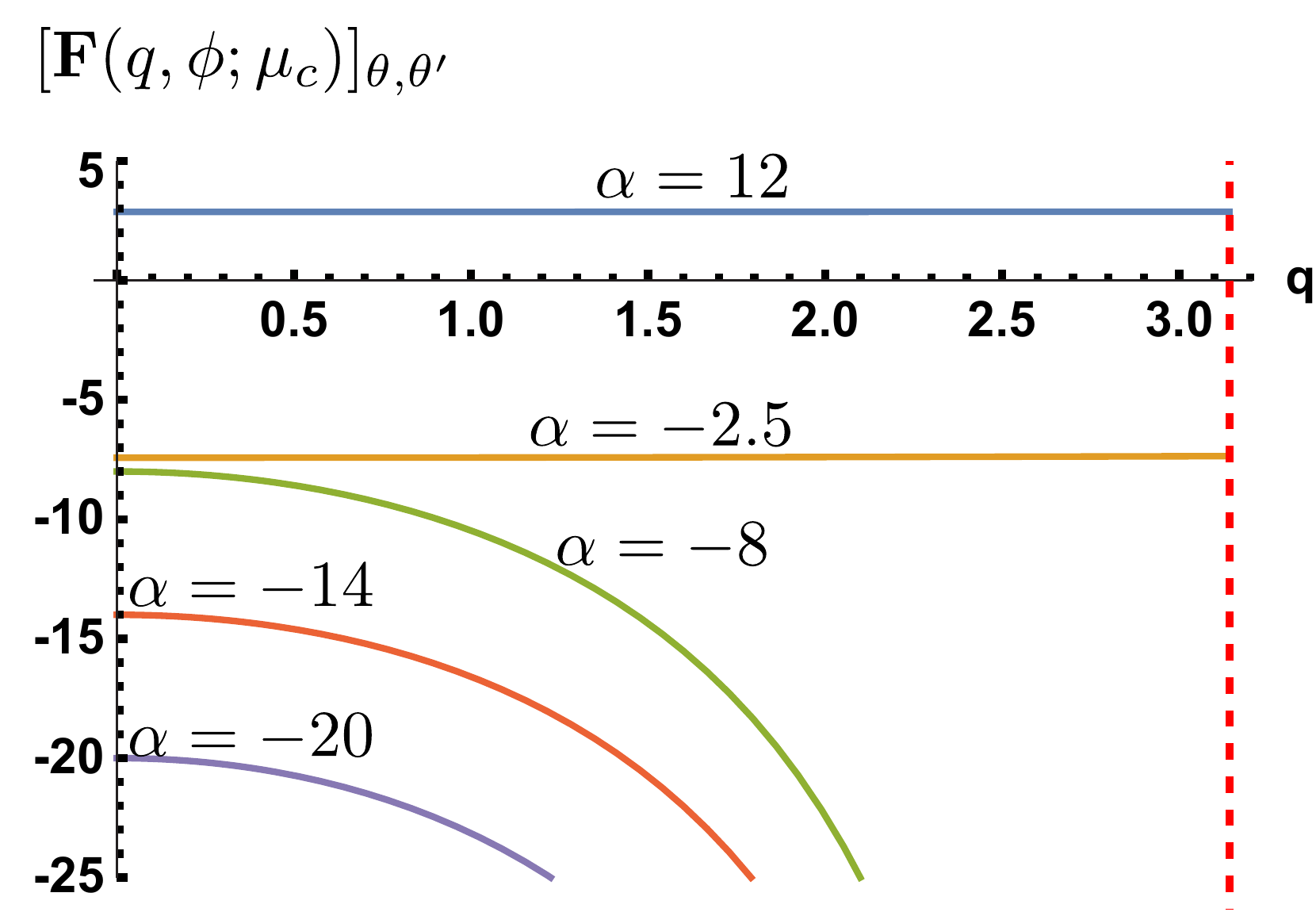}
    \caption{
    The profile of the coupling function that emerges in the $\mu \rightarrow \mu_c$ limit from the UV coupling function 
    $[{\bf F}(q,\phi;\Lambda)]_{\theta,\theta'} =\alpha$
    for different values of $\alpha$.
    For stable Fermi liquids,
    with $\alpha>-4\pi v_F \sqrt{1+v_F^2 \frac{q^2}{\Lambda^2}}$,
    $\mu_c=0$.
    For unstable cases,
    the critical energy scales are given by
    $\mu_c=0.580$, 
    $\mu_c=0.995$,  
    $\mu_c=1.288$ 
    for
    $\alpha=-8$, 
    $\alpha=-14$,
    $\alpha=-20$, respectively
    for the choice of 
    $v_F=0.3$ and $\Lambda=10$.
    \label{fig:zs_IR_swave}}
\end{figure}

Let us first consider the angle independent UV coupling function,
$[{\bf F}(q,\phi;\Lambda)]_{\theta,\theta'} =\alpha$.
For $\alpha > - 4\pi v_F $, 
eigenvalues of the coupling function
remain finite at all energy scales.
The coupling function that emerges in the $\mu \rightarrow 0$ limit represents 
the scale invariant Fermi liquid fixed point.
The renormalization group flow changes qualitatively 
for sufficiently attractive interaction with
$\alpha \leq  - 4\pi v_F $.
For $\alpha \leq  - 4\pi v_F $, 
the coupling function 
 at non-zero momenta can diverge at low energies.
As $\mu$ is lowered,
the divergence arises first at the momentum where the energy of particle-hole pair is peaked.
This is because the quantum correction in the particle-hole channel vanishes in the strict forward scattering limit
and increases with increasing energy of particle-hole pair.
The precise momentum at which the divergence arises at the highest energy depends on the full band structure.
To be concrete,
here we consider the simple case where the particle-hole energy 
 dispersion is well approximated by the linear dispersion 
 before it is peaked at $q_c$ and bends down at larger momenta.
 In this case, 
 \eq{eq:Fqphimu} holds upto $q \sim q_c$ and the coupling function diverges at $q=q_c$ as $\mu$ approaches a critical energy scale $\mu_c$.
The divergence of the four-fermion coupling at a non-zero $q$
can potentially represent a charge or spin density wave instability,
depending on whether the divergence is in
the spin singlet or the triplet  channel\cite{PhysRevB.70.155114}.
If the divergence arises 
at $q_c \gg \mu_c$, however,
one needs to be careful in interpreting it as a sign of instability
because
the low-energy effective theory description is not valid at $q$ larger than $\mu$.
With this cautionary remark,
we show the evolution of coupling in the deep attractive region in Fig.~\ref{fig:zs_RG_flow_Instability}.
In Fig.~\ref{fig:zs_IR_swave}, 
we show the evolution of $[\tilde{\bf F}(\tilde{q},\phi;\mu_c)]_{\theta,\theta'}$ 
as the strength of the UV coupling is tuned.

\subsubsection{the d-wave channel}

\begin{figure}[h]
    \centering
    \includegraphics[width=.9\textwidth]{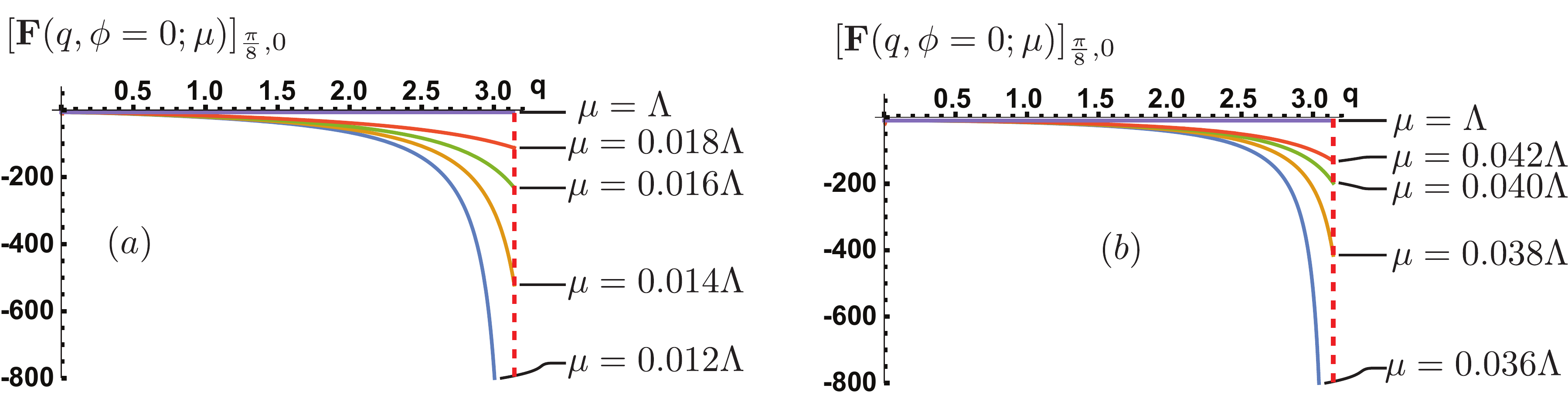}
    \caption{
        RG flow of 
$[{\bf F}({q},\phi=0;\mu)]_{\frac{\pi}{8},0}$ 
   for an attractive UV coupling function 
 $[{\bf F}(q,\phi;\Lambda)]_{\theta,\theta'} =2\alpha \cos [2(\theta-\theta')]$
    with 
    (a) $\alpha=-(4\pi v_F+1)$ 
    and
    (b) $\alpha=-(4\pi v_F+3)$. 
   The coupling function diverges at the momentum cutoff $q_c=\pi$
   at $\mu_c=0.122$ 
   and $\mu_c=0.362$, 
   respectively.
    We use $\Lambda=10$ and $v_F=0.3$.
     }
    \label{fig:evolve_f_dwave_instability}
\end{figure}
\begin{figure}[h]
    \centering
    \includegraphics[width=.9\textwidth]{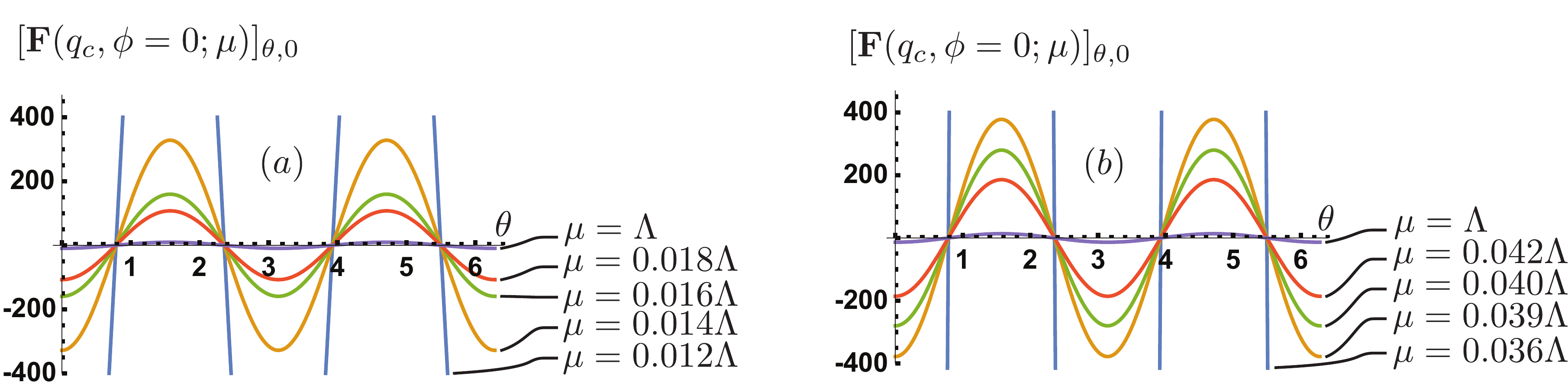}
    \caption{
    $[{\bf F}({q},\phi=0;\mu)]_{\theta,0}$ plotted as a function of $\theta$ at ${q}_c$ 
    at different values of $\mu$ 
    for
     (a) $\alpha=-(4\pi v_F+1)$ and 
     (b) $\alpha=-(4\pi v_F+3)$.
     The critical energy scales correspond to $\mu_c=0.122$ and $\mu_c=0.362$, respectively.
The same UV coupling function and parameters are used as in Fig.         \ref{fig:evolve_f_dwave_instability}.
    }
    \label{fig:evolve_f_dwave_theta_instability}
\end{figure}

Next, let us consider the case where 
the UV coupling is attractive in the d-wave channel with
$[{\bf F}(q,\phi;\infty)]_{\theta,\theta'}=2\alpha\cos [2(\theta-\theta')]$. 
In Fig.~\ref{fig:evolve_f_dwave_instability}, 
we plot the evolution of $[{\bf F}({q},\phi=0;\mu)]_{\frac{\pi}{8},0}$ as $\mu$ is lowered. 
For $\alpha \leq \alpha_c=-4\pi v_F$,
the coupling function 
at $q=q_c$   diverges 
as a critical energy scale 
$\mu_c$ is approached.
Fig.~\ref{fig:evolve_f_dwave_theta_instability}
shows the angular dependence of 
$[{\bf F}({q}_c,\phi=0;\mu)]_{\theta,0}$ 
at different $\mu/\Lambda$. 
A large attractive interaction
for particle-hole pairs
in the d-wave channel
with a non-zero momentum
promotes a distortion of the Fermi surface with a spatial modulation.
This corresponds to a bond density wave 
that causes a spatial modulation in the pattern of rotational symmetry breaking.

\begin{figure}[h]
    \centering
    \includegraphics[width=.43\textwidth]{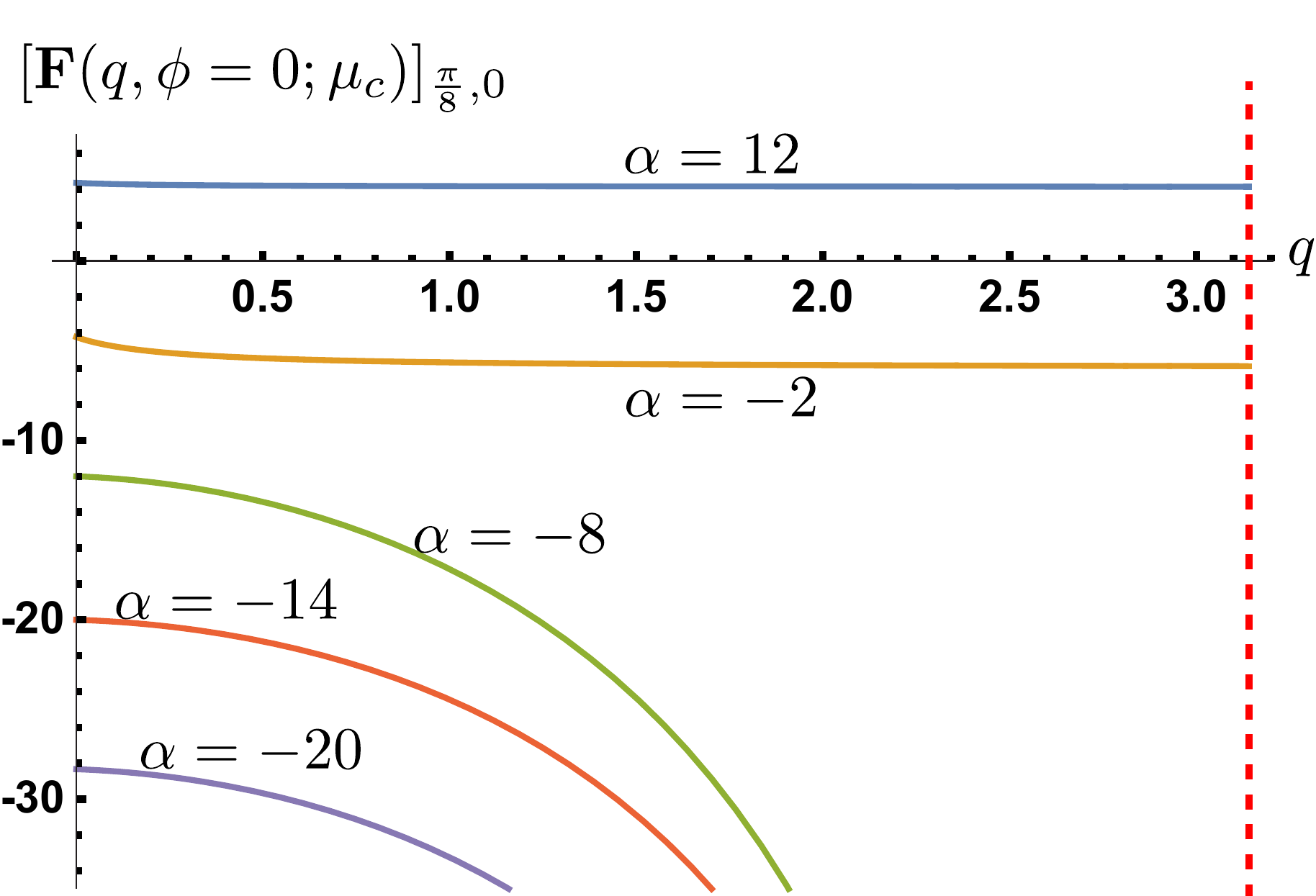}
    \caption{
        The profile of the coupling function that emerges in the $\mu \rightarrow \mu_c$ limit from the UV coupling function 
         $[{\bf F}(q,\phi;\Lambda)]_{\theta,\theta'} =2\alpha \cos [2(\theta-\theta')]$
    for different values of $\alpha$.
    $\mu_c=0$ for stable Fermi liquids with $\alpha>-4\pi v_F \sqrt{1+v_F^2 \frac{q^2}{\Lambda^2}}$.
    For unstable cases,
    the critical energy scales are given by
    $\mu_c=0.491$,
    $\mu_c=0.937$, 
    $\mu_c=1.243$ 
    for $\alpha=-8$, 
    $\alpha=-14$, 
    $\alpha=-20$, respectively for $v_F=0.3$ and $\Lambda=10$.
    }
    \label{fig:zs_IR_dwave}
\end{figure}

\begin{figure}[h]
    \centering
    \includegraphics[width=.9\textwidth]{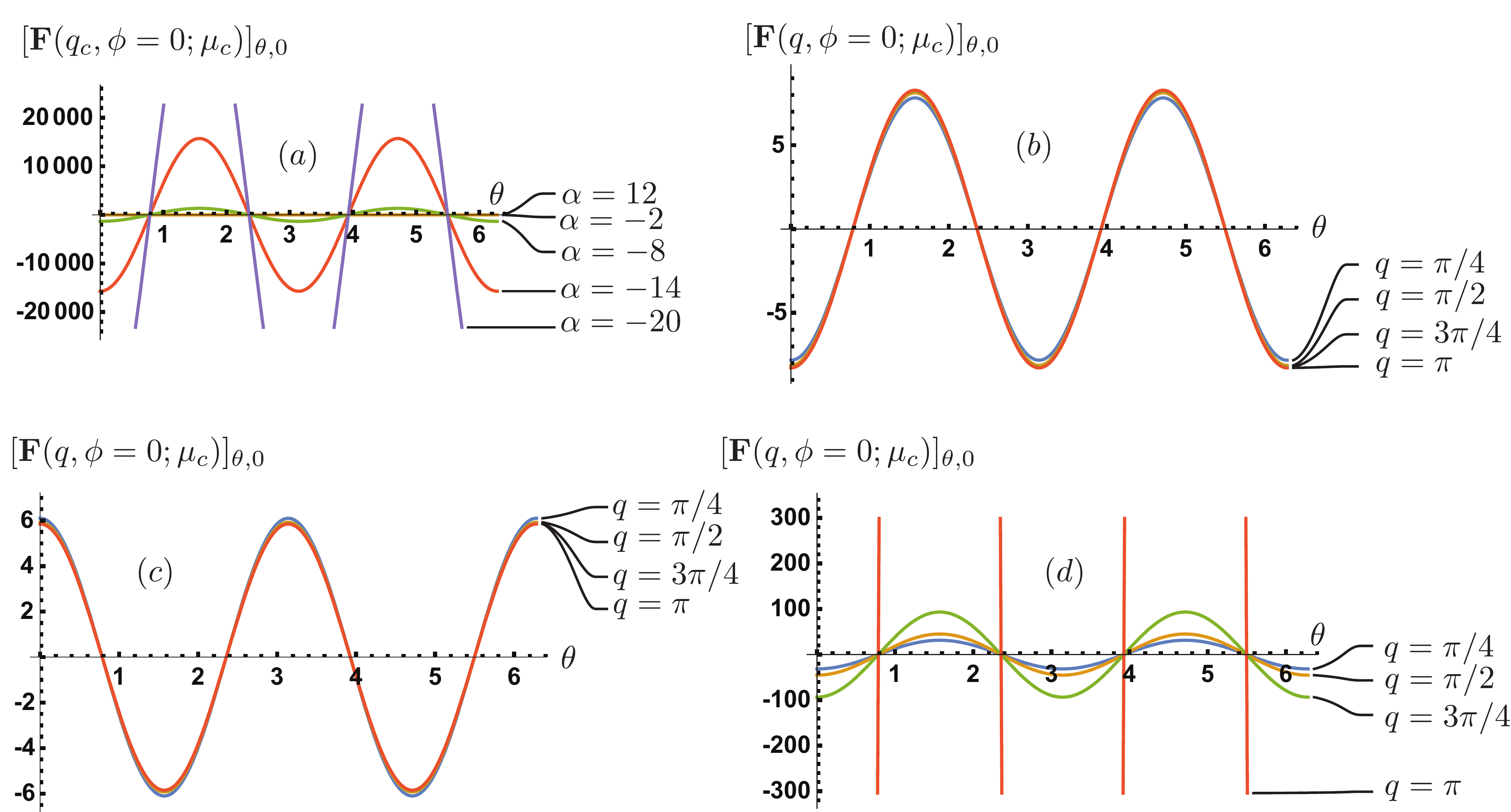}
    \caption{
 The coupling function at the critical energy scale
 plotted as a function of $\theta$ and $q$
 for the UV coupling function 
         $[{\bf F}(q,\phi;\Lambda)]_{\theta,\theta'} =2\alpha \cos [2(\theta-\theta')]$.
(a) The coupling function at 
 $\mu=\mu_c$ at $q=q_c$ for various choices of  $\alpha$.
(b) The coupling function 
 at $\mu=0$ for $\alpha=12$.
(c) The coupling function at $\mu=0$ for $\alpha=-2$.
(d) The coupling function 
at $\mu_c=0.937$ 
for various choices of momenta
for $\alpha=-14$.
$\Lambda=10$ and $v_F=0.3$ are used for all plots. 
    }
    \label{fig:zs_dwave_theta_IR}
\end{figure}

Fig.~\ref{fig:zs_dwave_theta_IR} shows how
$ [{\bf F}( {q},\phi=0;\mu_c)]_{\theta,0}$ evolves as $\alpha$ is varied.
For $\alpha > \alpha_c$, the coupling function converges to a 
fixed profile in the low-energy limit.
On the other hand, the amplitude of the coupling function grows without a bound as $\mu$  approaches $\mu_c$ for $\alpha \leq \alpha_c$.

\section{
Pairing channel
\label{sec:BCS}}

In this section, 
we discuss the renormalization group flow for the general pairing interaction that includes Cooper pairs with non-zero center of mass momentum.
The counter term from the one-loop vertex correction reads
\begin{eqnarray}
&& 
(A_1^\lambda )_{\theta-\varphi_\theta^S,\sigma_4;\theta+\varphi_\theta^S+\pi,\sigma_3}^{\theta'-\varphi_{\theta'}^S,\sigma_1;\theta'+\varphi_{\theta'}^S+\pi,\sigma_2} 
(\lambda_1)_{\theta-\varphi_\theta^S,\sigma_4;\theta+\varphi_\theta^S+\pi,\sigma_3}^{\theta'-\varphi_{\theta'}^S,\sigma_1;\theta'+\varphi_{\theta'}^S+\pi,\sigma_2} 
=  \nn
&&
\frac{1}{2 \mu} 
\frac{k_F}{\mu} 
\int \frac{d\theta'' d \kappa'' }{(2\pi)^2} \frac{d\Omega''}{2\pi}
\sum_{\sigma',\sigma''}~ 
(\tilde \lambda_1)_{
\theta''-\varphi_{\theta''}^S,\sigma';
\theta''+\varphi_{\theta''}^S+\pi,\sigma''}^{
\theta'-\varphi_{\theta'}^S,\sigma_1;
\theta'+\varphi_{\theta'}^S+\pi,\sigma_2} 
(\tilde \lambda_1)_{\theta-\varphi_\theta^S,\sigma_4;\theta+\varphi_\theta^S+\pi,\sigma_3}^{
\theta''-\varphi_{\theta''}^S,\sigma';
\theta''+\varphi_{\theta''}^S+\pi,\sigma''}  \nn
&& \times
Re \left[
\frac{1}{-i\Omega''
+\frac{1}{2m}\left[2k_F\kappa''+(\kappa'')^2+ (k_F+\kappa'')Q \cos (\theta''-\Phi)+\frac{Q^2}{4}\right]}
\right.
\nonumber\\ && 
~~~~~ \left.
~~~~~ \times \frac{1}{
-i(\mu-\Omega'')
+\frac{1}{2m}\left[2k_F\kappa''+(\kappa'')^2- (k_F+\kappa'')Q \cos (\theta''-\Phi)+\frac{Q^2}{4}\right]} 
\right].
\end{eqnarray}
The factor of $k_F/\mu$ represents the extensive phase space available for virtual Cooper pairs in the loop.
We define a dimensionless coupling function 
that incorporates the phase space as
$[V_{\theta',\theta}(Q,\Phi)]^{\sigma_1,\sigma_2}_{\sigma_4,\sigma_3}
\equiv
\frac{k_F}{\mu} (\tilde \lambda_1)^{\theta'-\varphi_{\theta'}^S,\sigma_1; \theta'+\varphi_{\theta'}^S+\pi ,\sigma_2}_{\theta-\varphi_\theta^S,\sigma_4; \theta+\varphi_\theta^S+\pi ,\sigma_3}$.
The beta functional for this new coupling is given by
\begin{eqnarray}
\frac{d V^r_{\theta_1,\theta_2}(Q,\Phi) }{dl} 
= -\frac{1}{8\pi^2 v_F}   \int d\theta ~ V^r_{\theta_1,\theta}(Q,\Phi)
\left[\frac{\mu^2}{
\mu^2 +  v_F^2 Q^2 \cos^2( \theta-\Phi)}
\right]
~V^r_{\theta,\theta_2}(Q,\Phi).
\label{eq:beta_BCS}
\end{eqnarray}
Here, $r=+$ or $-$.
$V^+_{\theta_1,\theta_2}(Q,\Phi)$
and
$V^-_{\theta_1,\theta_2}(Q,\Phi)$
represent the pairing interactions
in the spin triplet 
and singlet channels, respectively,
\bqa
[V_{\theta',\theta}(Q,\Phi)]^{\sigma_1,\sigma_2}_{\sigma_4,\sigma_3}
=
\mathcal{S}_{\sigma_4,\sigma_3}^{\sigma_1,\sigma_2}  
V^+_{\theta_1,\theta_2}(Q,\Phi)
+
\mathcal{A}_{\sigma_4,\sigma_3}^{\sigma_1,\sigma_2}  
V^-_{\theta_1,\theta_2}(Q,\Phi),
\eqa
where
$\mathcal{S}_{\sigma_4,\sigma_3}^{\sigma_1,\sigma_2} =\frac{1}{2}(\delta^{\sigma_1}_{\sigma_3}\delta^{\sigma_2}_{\sigma_4}+\delta^{\sigma_1}_{\sigma_4}\delta^{\sigma_2}_{\sigma_3} )$ and
$\mathcal{A}_{\sigma_4,\sigma_3}^{\sigma_1,\sigma_2} =
\frac{1}{2}(\delta^{\sigma_1}_{\sigma_4}\delta^{\sigma_2}_{\sigma_3} -\delta^{\sigma_1}_{\sigma_3}\delta^{\sigma_2}_{\sigma_4})$.
From now on, we focus on one spin channel
and omit the superscript $r$ in $V$.
For Cooper pairs with zero center of mass momentum, we set $Q=0$ to reproduce the well-known beta functional \cite{SHANKAR},
$
\frac{d V_{\theta_1,\theta_2}(0,0) }{dl} 
=-\frac{1}{8\pi^2 v_F} ~\int d\theta~ V_{\theta_1,\theta}(0,0)V_{\theta,\theta_2}(0,0)
$.

For general $Q$, the solution of 
 the beta functional is written as
\begin{eqnarray}
[{\bf V}(Q,\Phi;\mu)]_{\theta,\theta'}^{-1} &=& 
\frac{
\delta(\theta-\theta')
}{4 v_F} 
\log \left[
\frac{
\Lambda^2+v_F^2 Q^2 \cos^2(\theta-\Phi) }
{\mu^2+v_F^2 Q^2 \cos^2(\theta-\Phi) }
\right]
+ [{\bf V}(Q,\Phi;\Lambda)]_{\theta,\theta'}^{-1},
\label{eq:bcs_solution_Q_theta}
\end{eqnarray}
where 
$[{\bf V}(Q,\Phi;\Lambda)]_{\theta,\theta'}$
denotes the pairing interaction
at UV cutoff $\Lambda$.
The most important aspect of
\eq{eq:bcs_solution_Q_theta}
is the logarithmic singularity
that is present at $Q=0$ and $\mu=0$.
If $[{\bf V}(Q,\Phi;\Lambda)]_{\theta,\theta'}$
is repulsive in all angular momentum channels,
the coupling function flows to zero logarithmically.
On the other hand, 
the coupling function at $Q=0$ diverges 
at a critical energy scale
if $[{\bf V}(Q,\Phi;\Lambda)]_{\theta,\theta'}$
has any channel with a negative eigenvalue.
This is the well-known BCS instability.
Here, we focus on the scaling behaviour of the coupling function at non-zero $Q$.
To remove the logarithmic divergence, we consider 
the difference of 
\eq{eq:bcs_solution_Q_theta} 
at two momenta, $Q$ and $Q_*$.
In the $\mu \rightarrow 0$ limit with fixed $\tilde Q = Q/\mu$ and $\tilde Q_* = Q_*/\mu$,
the difference becomes
\begin{eqnarray}
\lim_{\mu \rightarrow 0}
\left(
[\tilde{\bf V}(\tilde{Q},\Phi;\mu)]^{-1}_{\theta,\theta'} 
- [\tilde{\bf V}(\tilde{Q}_*,\Phi;\mu)]^{-1}_{\theta,\theta'}
\right)
&=&
\frac{
\delta(\theta-\theta')
}{4 v_F} 
\log \left[
\frac{
1+v_F^2 \tilde Q_*^2 \cos^2(\theta-\Phi) }
{1+v_F^2 \tilde Q^2 \cos^2(\theta-\Phi) }
\right]
\label{eq:diffV}
\end{eqnarray}
where 
$[\tilde{\bf V}(\tilde{Q},\Phi;\mu)]_{\theta,\theta'} 
=
[{\bf V}(
\mu \tilde{Q},\Phi;\mu)]_{\theta,\theta'}$.

\begin{figure}[h]
    \centering
    \includegraphics[width=.99\textwidth]{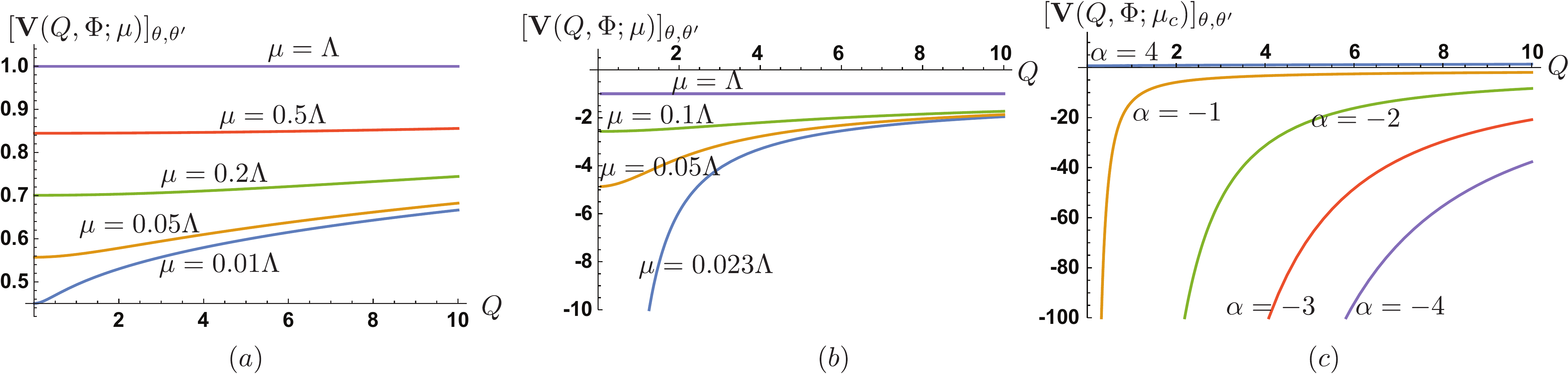}
    \caption{ 
    The RG evolution of $[{\bf V}({Q},\phi;\mu)]_{\theta,\theta'}$ 
    for the UV coupling in the s-wave channel with strength
    (a) $V_0(Q;\Lambda)=1$ and (b)  $V_0(Q;\Lambda)=-1$ with $\mu_c=0.023$. 
    (c) The coupling function that emerges in the 
    $\mu \rightarrow \mu_c$ limit for different UV couplings.
    For repulsive UV couplings 
 with $V_0(Q;\Lambda)>0$, 
the coupling function vanishes in the IR limit ($\mu_c=0$). 
    For attractive couplings with
   $V_0(Q;\Lambda)=\alpha_V=-1,-2,-3,-4$, the pairing interaction $Q=0$ diverges at $\mu_c=0.23, 1.51,2.84,3.89$, respectively. 
   We use $\Lambda=10$ and $v_F=0.3$.
        }
    \label{fig:bcs_RG_flow}
\end{figure}

Now, let us consider a simple case the UV coupling function is non-zero only in the s-wave spin-singlet channel.
The beta functional for the coupling in the s-wave channel becomes
\begin{eqnarray}
\frac{dV_0(Q;\mu)}{dl} &=&  
-\frac{1}{8\pi^2 v_F}  
V_0^2(Q;\mu) 
\int d\theta ~  
\frac{\mu^2}{
\mu^2 + v_F^2 Q^2 \cos^2(\theta)
},
\label{eq:BCS_beta_s_wave}
\end{eqnarray}
where $V_0$ represents the coupling in the s-wave channel.
The solution is written as
\begin{eqnarray}
V_0(Q;\mu) = \frac{1}{\frac{1}{8\pi v_F } \log \left[\frac{\sqrt{v_F^2Q^2+\Lambda^2}+\Lambda}{\sqrt{v_F^2Q^2+\Lambda^2}-\Lambda}\frac{\sqrt{v_F^2Q^2+\mu^2}-\mu}{\sqrt{v_F^2Q^2+\mu^2}+\mu} \right]+[V_0(Q;\Lambda)]^{-1}},
\label{eq:V_0}
\end{eqnarray}
where $V_{0}(Q;\Lambda)$ is the s-wave coupling
defined at UV cutoff scale $\Lambda$.
In Fig.~\ref{fig:bcs_RG_flow}, we plot the evolution of the coupling functions for different choices of the UV coupling in the s-wave channel.
In this example,
the scale invariance 
is expressed as
\begin{eqnarray}
\lim_{\mu \rightarrow 0}
\tilde{V}_0(\tilde{Q};\mu) = 
    \frac{1}{[\tilde{V}_0(\tilde{Q}_*)]^{-1} +
\frac{1}{8\pi v_F } \log \left[\frac{\sqrt{v_F^2 \tilde Q_*^2+1}+1}{\sqrt{v_F^2 \tilde Q_*^2+1}-1}
\frac{\sqrt{v_F^2 \tilde Q^2+1}-1}{\sqrt{v_F^2 \tilde Q^2+1}+1} \right]
     }
\end{eqnarray}
in the $\mu \rightarrow 0$ limit 
with fixed $\tilde Q$ and $\tilde Q_*$.

\section{Experimental consequences \label{sec:experiment}}

In this section,  we discuss how the universal coupling functions manifest themselves in physical observables.
In particular, we show that our local effective field theory contain all dynamical information for low-energy collective modes.
This is in contrast to 
 the fact that Landau's fixed point theory 
and the earlier RG schemes that do not keep track of the universal momentum-dependence of the coupling functions  can not capture the collective modes and
one has to resort to more microscopic theories to describe them.
Below, we examine the bosonic collective modes with charge $0$ and $2$, respectively.

\subsection{
Collective mode with charge $0$ :
density-density correlation function}

Collective modes with zero charge and spin can be probed through the density-density correlation function.
In Matsubara frequency, 
it is written as
\begin{eqnarray}
    \chi_{ph}(\vec{q},i\omega_n) 
    &=&\frac{1}{2} \sum_{\sigma,\sigma'} \int_0^\infty d\tau ~e^{i\omega_n \tau } \int \frac{d^2 \vec{k}}{(2\pi)^2} \frac{d^2\vec{p}}{(2\pi)^2} ~\langle T_\tau \psi_{\vec{k}+\frac{\vec{q}}{2},\sigma}^\dag (
    \tau) \psi_{\vec{k}-\frac{\vec{q}}{2},\sigma} (\tau) \psi_{\vec{p}-\frac{\vec{q}}{2},\sigma'}^\dag (
    0) \psi_{\vec{p}+\frac{\vec{q}}{2},\sigma'} (0) \rangle.  \nn
\end{eqnarray}
It can be expressed as 
$\chi_{ph}=\chi^{(0)}_{ph}+\chi^{(1)}_{ph}$,
where
$    \chi^{(0)}_{ph}(\vec{q},i\omega_n) 
    =\frac{k_F}{4\pi v_F} \left[1-\frac{1}{\sqrt{1+\frac{v_F^2q^2}{\omega^2_n}}}\right]$
is the free electron contribution,
and
$\chi^{(1)}_{ph}$ is the interaction part.
The interacting part is obtained by connecting the electron propagators to the one-particle irreducible quartic vertex function and summing over the relative momenta of particle-hole pairs as is shown below.
\begin{eqnarray}
\Gamma^{(4)} &=& 
\begin{tikzpicture}[baseline={([yshift=-4pt]current bounding box.center)}]
\coordinate (v1) at (-20pt,10pt);
\coordinate (v2) at (20pt,10pt);
\coordinate (v3) at (-20pt,-10pt);
\coordinate (v4) at (20pt,-10pt);
\coordinate (v5) at (0pt,0pt);
\draw[thick,postaction={mid arrow=red} ] (v1) to  (v5);
\draw[thick,postaction={mid arrow=red} ] (v5) to  (v2);
\draw[thick,postaction={mid arrow=red} ] (v5) to  (v3);
\draw[thick,postaction={mid arrow=red} ] (v4) to  (v5);
\node at (v5)[rectangle,fill,inner sep=3pt]{};
\end{tikzpicture} , \quad \quad \quad
\chi_{ph}
=
\begin{tikzpicture}[baseline={([yshift=-4pt]current bounding box.center)}]
\coordinate (v1) at (-20pt,0pt);
\coordinate (v2) at (20pt,0pt);
\draw[thick,postaction={mid arrow=red} ] (v1) to [bend left=40] (v2);
\draw[thick,postaction={mid arrow=red} ] (v2) to [bend left=40] (v1);
\node at (v1)[circle,fill,inner sep=1pt]{};
\node at (v2)[circle,fill,inner sep=1pt]{};
\end{tikzpicture} \quad 
+
\quad
\begin{tikzpicture}[baseline={([yshift=-4pt]current bounding box.center)}]
\coordinate (v1) at (-20pt,0pt);
\coordinate (v2) at (20pt,0pt);
\coordinate (v3) at (60pt,0pt);
\draw[thick,postaction={mid arrow=red} ] (v1) to [bend left=40] (v2);
\draw[thick,postaction={mid arrow=red} ] (v2) to [bend left=40] (v1);
\draw[thick,postaction={mid arrow=red} ] (v2) to [bend left=40] (v3);
\draw[thick,postaction={mid arrow=red} ] (v3) to [bend left=40] (v2);
\node at (v1)[circle,fill,inner sep=1pt]{};
\node at (v2)[rectangle,fill,inner sep=3pt]{};
\node at (v3)[circle,fill,inner sep=1pt]{};
\end{tikzpicture}
\label{eq:Gamma4chi}
\end{eqnarray}
The non-trivial dynamics is encoded in the energy and momentum dependence of the vertex function.
In our description, the coupling functions that depend on the momenta along the Fermi surface 
and the renormalization group energy scale contain that information 
through Eqs. \eqref{eq:RG4}-\eqref{eq:RG5}.
In the limit that the momentum carried by a particle-hole pair is small,
the nearly forward scattering processes are most important.
In this case, we can obtain the interacting part of the correlation function using our renormalized coupling function as 
\begin{eqnarray}
    \chi^{(1)}_{ph}(\vec{q},i\omega_n) &=& - \frac{1}{2\beta^2}\sum_{\substack{\sigma,\sigma'\\ \sigma_{1}, \sigma_{2} ,\sigma_{3} ,\sigma_{4}}}\sum_{ i\nu_n,i\nu'_n} \int \frac{d^2 \vec{k}}{(2\pi)^2} \frac{d^2 \vec{p}}{(2\pi)^2} ~G_{\sigma}(\vec{k}+\frac{\vec{q}}{2},i\omega_n+i\nu_n) G_{\sigma}(\vec{k}-\frac{\vec{q}}{2},i\nu_n) \nonumber\\
    &\times& \delta_{\sigma_1}^{\sigma} \delta_{\sigma_4}^{\sigma} \delta_{\sigma_2}^{\sigma'}\delta_{\sigma_3}^{\sigma'}   (\lambda_0)^{\vec{k}+\frac{\vec{q}}{2},\sigma_1;\vec{p}-\frac{\vec{q}}{2},\sigma_2}_{\vec{k}-\frac{\vec{q}}{2},\sigma_4;\vec{p}+\frac{\vec{q}}{2},\sigma_3} G_{\sigma'}(\vec{p}+\frac{\vec{q}}{2},i\omega_n+i\nu'_n) G_{\sigma'}(\vec{p}-\frac{\vec{q}}{2},i\nu'_n) \nonumber\\
    &=&  -\frac{k_F}{(4\pi v_F)^2}\int \frac{d \theta}{2\pi} \frac{d \theta'}{2\pi} ~ \left[ \frac{v_F q \cos \theta}{-i\omega_n+v_Fq\cos \theta }\right][{\bf F}(q,\phi;\omega_n)]_{\theta,\theta'}\left[ \frac{v_F q \cos \theta'}{-i\omega_n+v_Fq\cos \theta' }\right].
    \label{eq:PHchi1}
\end{eqnarray}

Below, we focus on the simplest case where the interaction is isotropic. 
For an interaction that is independent of angles at a UV scale $\Lambda$,
the renormalized coupling function at energy scale $\mu$ is given by
\eq{eq:FwithLambda}.
In this case, the density-density correlation function is obtained to be
\begin{eqnarray}
    \chi_{ph}(\vec{q},i\omega_n) &=& \chi^{(0)}_{ph}(\vec{q},i\omega_n) + \chi^{(1)}_{ph}(\vec{q},i\omega_n) \nonumber\\
    &=&\frac{k_F}{4\pi v_F}\left\{ \frac{\alpha}{4\pi v_F}\left[\frac{1}{\sqrt{1+\frac{v_F^2q^2}{\Lambda^2}}}-1\right]+1 \right\}~\frac{\left[1-\frac{1}{\sqrt{1+\frac{v_F^2q^2}{\omega_n^2}}}\right]}{\frac{\alpha}{4\pi v_F}\left[\frac{1}{\sqrt{1+\frac{v_F^2q^2}{\Lambda^2}}}-\frac{1}{\sqrt{1+\frac{v_F^2q^2}{\omega_n^2}}}\right]+1}.
\end{eqnarray}
The retarded Green's function is obtained through the analytical continuation $i\omega_n \rightarrow \omega +i\delta$, 
\begin{eqnarray}
    \chi_{ph}(\vec{q},\omega) &=& \frac{k_F}{4\pi v_F}\left\{ \frac{\alpha}{4\pi v_F}\left[\frac{1}{\sqrt{1+\frac{v_F^2q^2}{\Lambda^2}}}-1\right]+1 \right\}~\frac{\left[1-\frac{1}{\sqrt{1-\frac{v_F^2q^2}{\omega^2}}}\right]}{\frac{\alpha}{4\pi v_F}\left[\frac{1}{\sqrt{1+\frac{v_F^2q^2}{\Lambda^2}}}-\frac{1}{\sqrt{1-\frac{v_F^2q^2}{\omega^2}}}\right]+1}.
\end{eqnarray}
Its imaginary part, 
which describes the spectral function for the neutral bosonic excitations,
has a branch cut at $\omega = v_Fq$ associated with the particle-hole continuum, 
\begin{eqnarray}
       \textrm{Im}\chi_{ph}(\vec{q},\omega) =
      \left\{ 
      \begin{array}{ccc}
       \frac{k_F}{4\pi v_F} ( \mathcal{A} -1)^2\left[\mathcal{A}^2-1\right]^{-3/2}
       \delta\left[\frac{\omega}{v_Fq}- \frac{\mathcal{A}}{\sqrt{\mathcal{A}^2-1}}\right]
      & \mbox{for} & \omega > v_F q \textrm{~~and~~} \alpha>0

       \\
      \frac{k_F}{4\pi v_F} (\mathcal{A}-1)^2~\sqrt{\frac{v_F^2q^2}{\omega^2}-1}~\left\{\mathcal{A}^2\left[\frac{v_F^2q^2}{\omega^2}-1\right]+1\right\}^{-1}
      & &\mbox{for~~}  \omega < v_F q
       \end{array}\right.,
       \label{eq:Imphiph}
\end{eqnarray}
where $\mathcal{A}=\frac{4\pi v_F}{\alpha}+\frac{1}{\sqrt{1+\frac{v_F^2q^2}{\Lambda^2}}}$.

\begin{figure}[h]
    \centering
    \includegraphics[width=.45\textwidth]{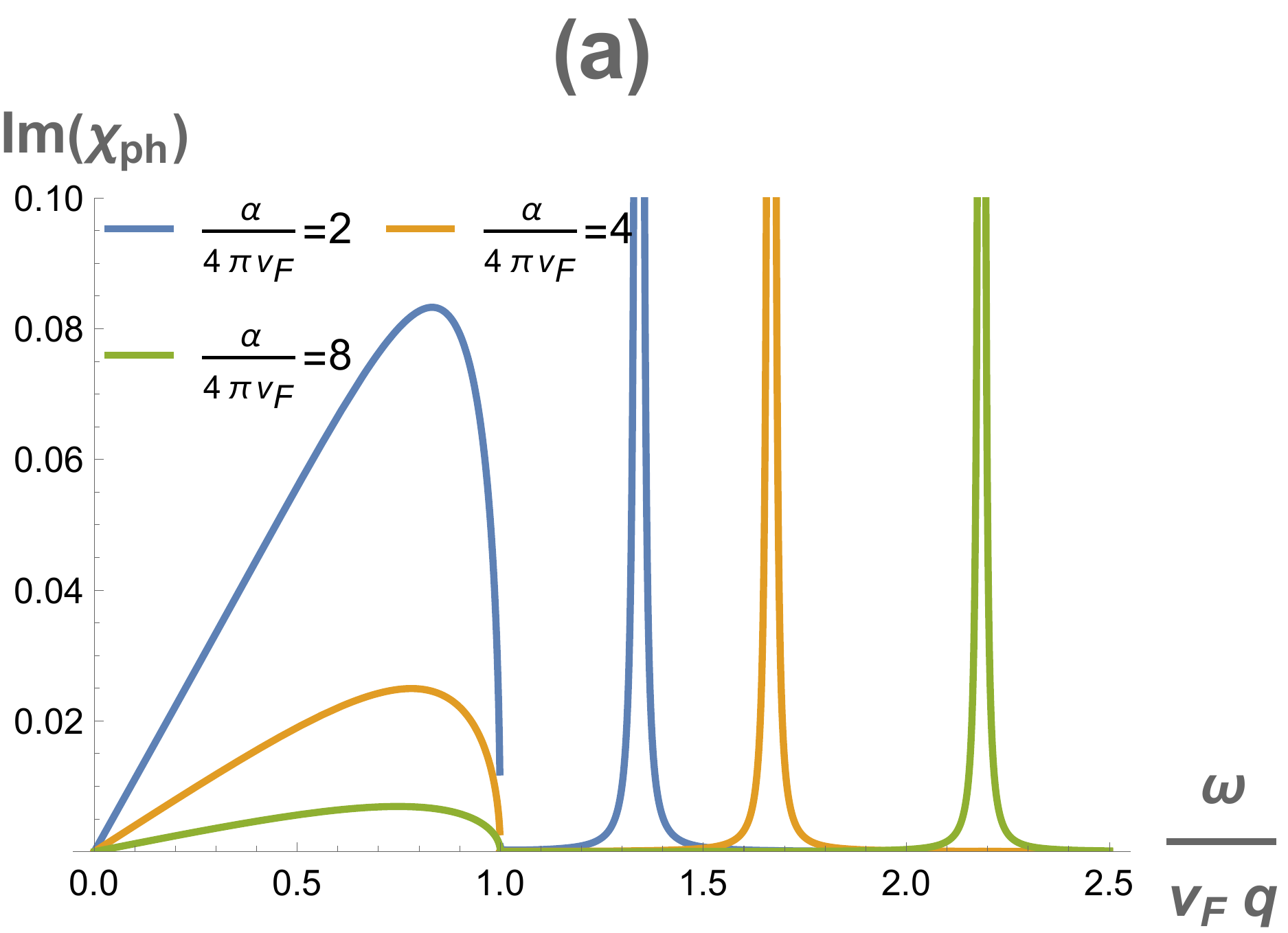}
    \includegraphics[width=.45\textwidth]{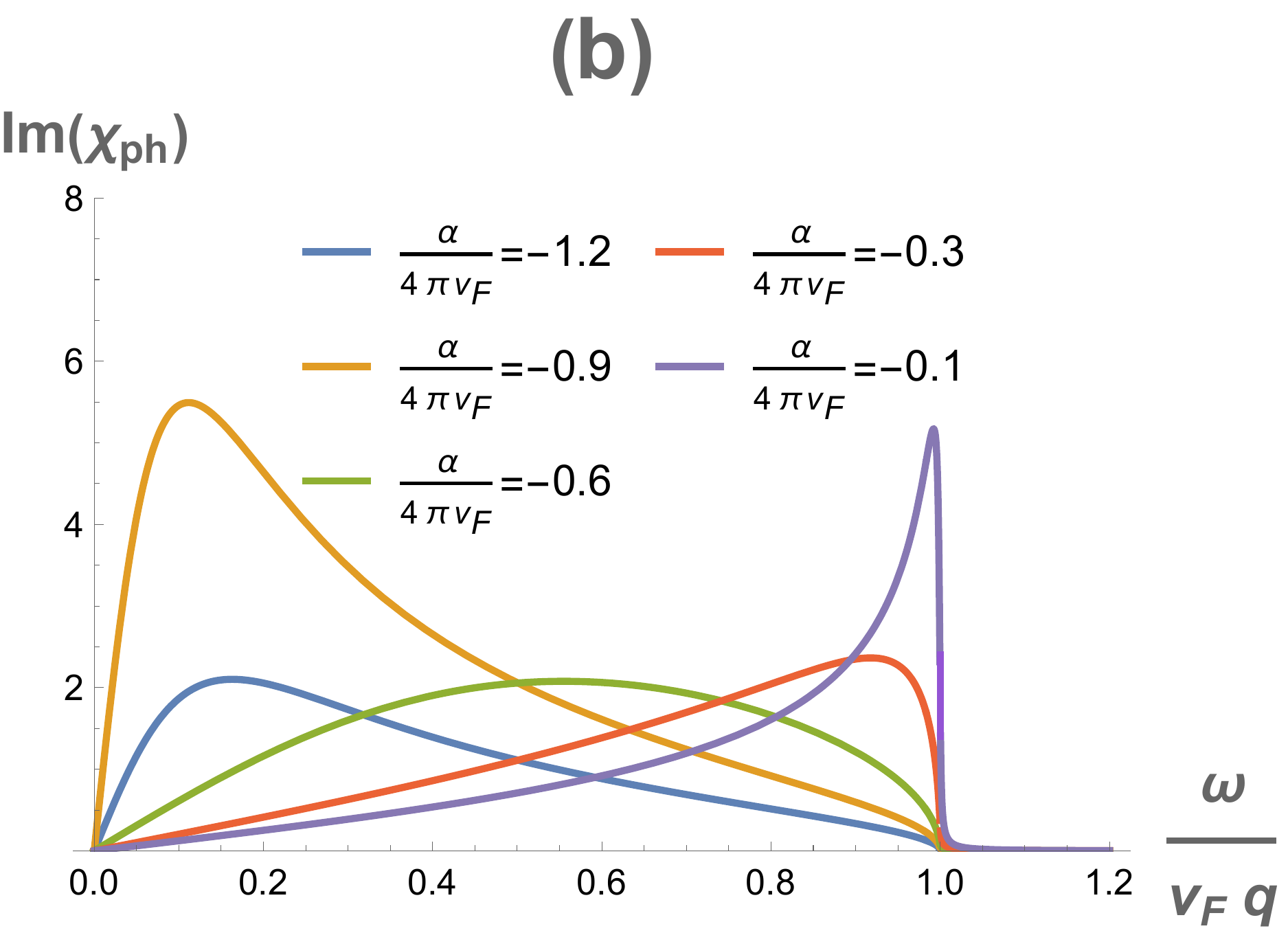}
    \caption{$\textrm{Im}\chi_{ph}(\vec{q},\omega)$ as a function of $\frac{\omega}{v_Fq} $ (horizontal axis) at various positive $\frac{\alpha}{4\pi v_F}$ and (b) negative $\frac{\alpha}{4\pi v_F}$. The zero sound modes correspond to the sharp peaks at $\frac{\omega}{v_Fq} \geq 1$. The damped modes corresponds to the broad peaks at $\frac{\omega}{v_Fq} < 1$. $\Lambda=40$ and $\delta=0.0001$.}
    \label{fig:spectral_ph}
\end{figure}

For a repulsive interaction ($\alpha>0$), the spectral function supports a delta-function peak outside the particle-hole continuum, 
\begin{eqnarray}
    \frac{\omega}{v_F q} =\frac{\mathcal{A}}{\sqrt{\mathcal{A}^2-1}}=\frac{1+\frac{\alpha}{4\pi v_F}\frac{1}{\sqrt{1+\frac{v_F^2q^2}{\Lambda^2}}}}{\sqrt{1+(\frac{\alpha}{4\pi v_F})^2\left[\frac{1}{1+\frac{v_F^2q^2}{\Lambda^2}}-1\right]+2\frac{\alpha}{4\pi v_F}\frac{1}{\sqrt{1+\frac{v_F^2q^2}{\Lambda^2}}}}}.
\end{eqnarray}
In the $\Lambda \rightarrow \infty$ limit, the dispersion of the collective mode can be written in a closed form as $\frac{\omega}{v_F q} =\frac{1+\frac{\alpha}{4\pi v_F}}{\sqrt{1+\frac{2\alpha}{4\pi v_F}}}$.
This corresponds to the zero sound mode.
The spectral weight of the mode increases linearly in the forward scattering amplitude in the weak coupling limit.
Besides the zero sound mode,
there also exists a broad peak inside the particle-hole continuum. 
The dispersion of the incoherent mode becomes
\begin{eqnarray}
\frac{\omega}{v_Fq}=\frac{\left(1+\frac{\alpha}{4\pi v_F}\right)}{ \sqrt{1+2\frac{\alpha}{4\pi v_F} \left(1+\frac{\alpha}{4\pi v_F}\right) }} < 1
    \label{eq:peak_Imchi}
\end{eqnarray}
in the $\Lambda \rightarrow \infty$ limit.
In the strong coupling limit,
the velocity of this mode 
 becomes $\frac{\omega}{v_F q} = \frac{1}{\sqrt{2}}$.
The height of the peak is 
$    \textrm{Max}[\textrm{Im}\chi] =   \frac{k_F}{4\pi v_F}\frac{1}{ 2\left(\frac{|\alpha|}{4\pi v_F}\right)\left(1+\frac{\alpha}{4\pi v_F}\right) } 
$
and the width at half maximum 
 is given by
\begin{eqnarray}
    \frac{\Delta \omega}{v_F q} = 
    \sqrt{\frac{(1+\frac{\alpha}{4\pi v_F})^2}{1+2\frac{\alpha}{4\pi v_F}+ 4( 2 - \sqrt{3})(\frac{\alpha}{4\pi v_F})^2}} - \sqrt{\frac{(1+\frac{\alpha}{4\pi v_F})^2}{1+2\frac{\alpha}{4\pi v_F}+ 4( 2 + \sqrt{3})(\frac{\alpha}{4\pi v_F})^2}}.
    \label{eq:width_first_sound}
\end{eqnarray}
At weak coupling, the width scales with the coupling as $\Delta \omega \approx 4\sqrt{3}(\frac{\alpha}{4\pi v_F})^2 v_F q$.
In the large $\alpha$ limit,  
the width becomes 
$\Delta \omega \approx \frac{1}{\sqrt{2}} v_F q$. 
In the presence of an attractive interaction,
the zero sound mode is no longer sharply defined as the speed of the mode goes below the Fermi velocity, and the zero sound mode merges with the incoherent mode\footnote{ The modes at $-\frac{1}{2}<\alpha<0$ are referred to as  Hidden modes\cite{Chubukov2019,klein2020hidden}}. The spectral function plotted  as a function of $\omega/(v_Fq)$ is shown in 
    Fig. \ref{fig:spectral_ph}
 for different interaction strengths.

\begin{figure}[h]
    \centering
    \includegraphics[width=.45\textwidth]{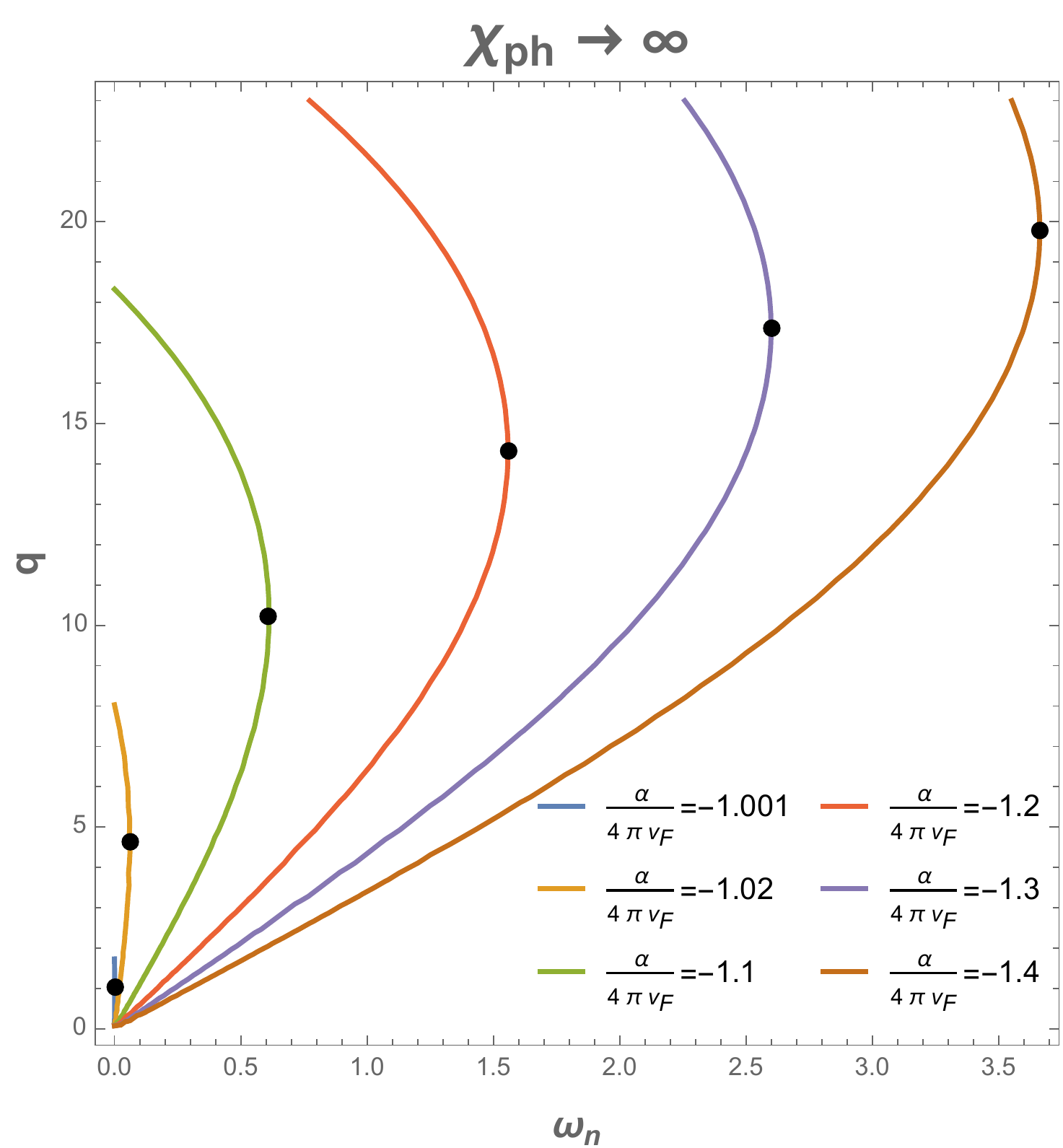}
    \caption{ 
    The location of poles 
 in the partice-hole spectral function as a function of $q$ and $\omega_n \equiv -i \omega$ for different choices of $\frac{\alpha}{4\pi v_F}$,
 where $\alpha$ denotes the bare interaction in the s-wave channel that is momentum independent at energy scale $\Lambda=40$.
 Black dots represent 
 $(\omega_{nc}(\alpha),q_c(\alpha))$ 
that correspond to  the pole with the largest imaginary frequency at each $\alpha$.
 }
    \label{fig:poles_sus_ph}
\end{figure}

For sufficiently strong attractive interaction with
$\alpha < \alpha_c= -4\pi v_F$, there exists a pole in the upper half plane of complex frequency at 
$\frac{\omega}{v_Fq}= i \frac{\mathcal{A}}{\sqrt{1-\mathcal{A}^2}}$. 
In the plane of $q$ and $\omega_n \equiv -i \omega$, 
the locations of the poles are shown 
for various values of the bare coupling in Fig.~\ref{fig:poles_sus_ph}. 
At $\alpha=\alpha_c$, the spectral function exhibits a pole only at $q = 0$ and $\omega_n=0$. 
In the channel in which the particle-hole pair carries spin one, this corresponds to the Stoner's instability associated with the ferromagnetic instability.
For more general UV interactions, the instability can arise in channels with non-zero angular momenta either in the spin singlet or triplet channels associated with
the Pomeranchuk instability. 
For $\alpha < \alpha_c$,  the spectral function exhibits a band of poles within 
$0 \leq \omega_n \leq 
\omega_{nc}(\alpha)$,
where $\omega_{nc}(\alpha)$ corresponds to the largest imaginary frequency
and
the momentum associate with the pole is determined through 
$\frac{\alpha}{4\pi v_F} = \frac{-1}{\frac{1}{\sqrt{1+\frac{v_F^2q^2}{\Lambda^2}}} -\frac{1}{\sqrt{1+\frac{v_F^2q^2}{\omega_n^2}}}}$.
If the imaginary frequency is lowered at a fixed $\alpha<\alpha_c$, the spectral function encounters the first pole at $\omega_{nc}(\alpha)$
with momentum,
\begin{eqnarray}
q_c(\alpha) 
=\Lambda \frac{\sqrt{2\sqrt{4(\frac{\alpha}{4\pi v_F})^2-1}
\left(\sqrt{4(\frac{\alpha}{4\pi v_F})^2-1}-\sqrt{3} \right)}}{(\sqrt{4(\frac{\alpha}{4\pi v_F})^2-1}+\sqrt{3})} \quad \textrm{ for }\quad \alpha<\alpha_c. 
\label{eq:qc}
\end{eqnarray} 

\begin{figure}[h]
    \centering
    \includegraphics[width=.7\textwidth]{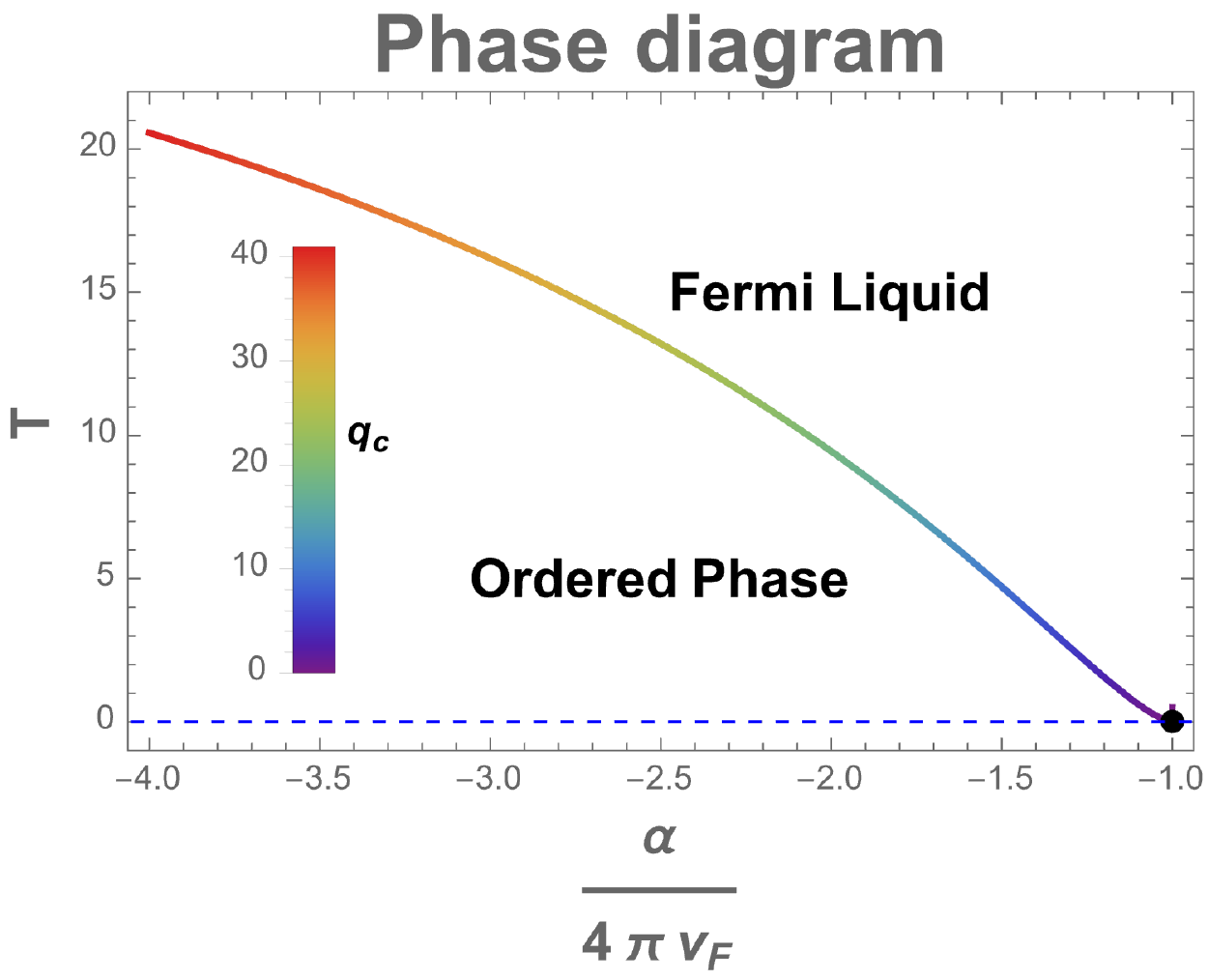}
    \caption{
    Phase diagram 
    of the two-dimensional Fermi liquids 
    in the plane of bare interaction $\alpha$
    and temperature $T$
    suggested from the one-loop renormalized coupling function.
   Here, we use $\omega_{nc}(\alpha)$ as the proxy for the transition temperature.
   The region above and below the curve represent the symmetric Fermi liquid and 
   a symmetry broken state, respectively.
   The colour changing along the phase boundary represents evolution of 
 the ordering wave-vector, $q_c(\alpha)$.
   As $\alpha$ decreases below $\alpha_c$,
   $q_c(\alpha)$ gradually increases from zero 
   following Eq.~\eqref{eq:qc}. 
   The black dot at $(-1.0,0.0)$ denotes the uniform Stoner's instability.
   }
   \label{fig:phase_diagram}
\end{figure}

The unstable modes 
 with non-zero momenta correspond to density-wave instabilities associated with charge, spin or nematicity depending on the quantum number of the particle-hole pair.
It is interesting to note that the strongest instability arises at $q \neq 0$ for $\alpha < \alpha_c$.
$\omega_{nc}(\alpha)$ can be viewed as a rough estimate of the transition temperature of the spontaneous symmetry breaking triggered by the instability.
In Fig.~\ref{fig:phase_diagram}, we illustrate the 
phase diagram that indicates the transition 
 temperature as a function of the interaction.

\subsection{
Collective mode with charge $2$ :
pair-pair correlation function}
The dynamics of 
 charge $2$ collective modes can be studied through the pair-pair correlation function.
The correlation function in the 
s-wave channel reads
\begin{eqnarray}
    &&(\chi_{pp})_{\sigma_4,\sigma_3}^{\sigma_1,\sigma_2}(\vec{Q},i\omega_n) 
    = \int_0^\infty d\tau ~e^{i\omega_n \tau } \int \frac{d^2 \vec{k}}{(2\pi)^2}\frac{d^2 \vec{k}'}{(2\pi)^2}   ~\langle T_\tau  \psi_{\vec{k}+\frac{\vec{Q}}{2},\sigma_4} (\tau)\psi_{-\vec{k}+\frac{\vec{Q}}{2},\sigma_3} (\tau) \psi_{-\vec{k}'+\frac{\vec{Q}}{2},\sigma_2}^\dag (0)\psi_{\vec{k}'+\frac{\vec{Q}}{2},\sigma_1}^\dag (0)  \rangle \nonumber\\
    &&=  \int \frac{d^2 \vec{k}}{(2\pi)^2}\frac{d^2 \vec{k}'}{(2\pi)^2}  \int^{\Lambda_{\omega}} \frac{d\omega'}{2\pi}\frac{d\Omega}{2\pi}\frac{d\Omega'}{2\pi} ~\langle  \psi_{\vec{k}+\frac{\vec{Q}}{2},\sigma_4} (\omega_n-\omega')  \psi_{-\vec{k}+\frac{\vec{Q}}{2},\sigma_3} (\omega') \psi_{-\vec{k}'+\frac{\vec{Q}}{2},\sigma_2}^\dag (\Omega)\psi_{\vec{k}'+\frac{\vec{Q}}{2},\sigma_1}^\dag (\Omega')  \rangle,
    \end{eqnarray}
    where $\vec Q$ and $\omega_n$ denote the center of mass momentum and energy of Cooper pairs.
It can be written as  
$\chi_{pp}=\chi^{(0)}_{pp}+\chi^{(1)}_{pp}$,
where $\chi^{(0)}_{pp}$ is the disconnected free-electron contribution
and $(\chi^{(1)}_{pp})_{\sigma_4,\sigma_3}^{\sigma_1,\sigma_2}(\vec{Q},i\omega_n)$
is the connected correlation function.
For $\omega_n > 0$, the free-electron part is written as 
\begin{eqnarray}
    (\chi^{(0)}_{pp})_{\sigma_4,\sigma_3}^{\sigma_1,\sigma_2}(\vec{Q},i\omega_n) 
    &=&-\mathcal{A}_{\sigma_4,\sigma_3}^{\sigma_1,\sigma_2} \frac{k_F}{\pi v_F}  \log \frac{\omega_n}{4\Lambda_\omega }\left(1+\sqrt{1+\frac{v_F^2Q^2}{\omega^2_n}}\right),
\end{eqnarray}
where
$\mathcal{A}_{\sigma_4,\sigma_3}^{\sigma_1,\sigma_2} =
\frac{1}{2}(\delta^{\sigma_1}_{\sigma_4}\delta^{\sigma_2}_{\sigma_3} -\delta^{\sigma_1}_{\sigma_3}\delta^{\sigma_2}_{\sigma_4})$.
Using Eqs. \eqref{eq:RG4}-\eqref{eq:RG5},
one can express the connected correlation function 
in terms of the renormalized coupling function as
\begin{eqnarray}
    (\chi^{(1)}_{pp})_{\sigma_4,\sigma_3}^{\sigma_1,\sigma_2}(\vec{Q},i\omega_n) 
    &=& -\frac{1}{2}\int \frac{d^2 \vec{k}_1}{(2\pi)^2} \frac{d^2 \vec{k}_2}{(2\pi)^2} \left[(\lambda_1)^{\vec{k}_2+\frac{\vec{Q}}{2},\sigma_1; -\vec{k}_2+\frac{\vec{Q}}{2},\sigma_2}_{\vec{k}_1+\frac{\vec{Q}}{2},\sigma_4; -\vec{k}_1+\frac{\vec{Q}}{2},\sigma_3} -(\lambda_1)^{\vec{k}_2+\frac{\vec{Q}}{2},\sigma_1; -\vec{k}_2+\frac{\vec{Q}}{2},\sigma_2}_{-\vec{k}_1+\frac{\vec{Q}}{2},\sigma_3; \vec{k}_1+\frac{\vec{Q}}{2},\sigma_4} \right]\int^{\Lambda_\omega} \frac{d\omega'}{2\pi}\frac{d\Omega}{2\pi}\nonumber\\
    &\times& G(\vec{k}_1+\frac{\vec{Q}}{2},\omega_n-\omega') G(-\vec{k}_1+\frac{\vec{Q}}{2},\omega') G (-\vec{k}_2+\frac{\vec{Q}}{2},\Omega) G (\vec{k}_2+\frac{\vec{Q}}{2},\omega_n-\Omega).
\end{eqnarray}
Let us consider an isotropic UV interaction that is momentum-independent :
$V_0(Q,\Lambda)=\alpha_V$, where
$\Lambda$ is the UV energy scale at which the bare coupling is defined.
In this case, the connected correlation function becomes
\begin{eqnarray}
    (\chi^{(1)}_{pp})_{\sigma_4,\sigma_3}^{\sigma_1,\sigma_2}(\vec{Q},i\omega_n) 
    &=& -\frac{1}{4k_F}\mathcal{A}_{\sigma_4,\sigma_3}^{\sigma_1,\sigma_2} V_0 (Q;\omega_n)  \left(\frac{k_F}{\pi v_F} \log \left[\frac{\omega_n }{4\Lambda_\omega}\left(1+\sqrt{1+\frac{v_F^2Q^2}{\omega^2_n}}\right)\right]\right)^2,
\end{eqnarray}
where $\Lambda_\omega$ is the frequency cutoff.
The precise value of the cutoff does not affect the low-energy physics. 
The correlation function 
in the spin singlet channel can be singled out as
$(\chi_{pp})_{\sigma_4,\sigma_3}^{\sigma_1,\sigma_2} (\vec{Q},i\omega_n)=\mathcal{A}_{\sigma_4,\sigma_3}^{\sigma_1,\sigma_2}\chi_{pp}^a (\vec{Q},i\omega_n)$, where 
\begin{eqnarray}
    \chi_{pp}^a( \vec{Q}, i\omega_n) &=& 
    \frac{k_F}{\pi v_F} \left[1+\frac{\alpha_V}{4\pi v_F}\log\frac{\Lambda}{2\Lambda_\omega} \right]\frac{4\pi v_F}{\alpha_V} \left\{ 1+\frac{1+\frac{\alpha_V}{4\pi v_F}\log \frac{\Lambda}{2\Lambda_\omega}}{\frac{\alpha_V}{4\pi v_F}\log \frac{\sqrt{v_F^2Q^2 +\omega_n^2}+\omega_n}{2\Lambda} - 1}\right\}.
    \label{eq:chiapp}
\end{eqnarray}
To obtain the pair Green's function in real frequency, 
we use the analytic continuation and 
define 
$\chi^a_{pp}(\vec{Q}, \omega + i 0^+)$.
In the complex plane of $\omega$,
$\chi^a_{pp}(\vec{Q},\omega+i0^+)$ has a pole at 
$\omega = i \omega_{Q}$ with 
\begin{eqnarray}
 \omega_{Q}= -e^{-\frac{4\pi v_F}{\alpha_V}} \frac{v_F^2Q^2}{4\Lambda} + e^{\frac{4\pi v_F}{\alpha_V}} \Lambda
\end{eqnarray}
and branch cuts slightly below the real axis with $|\omega| > v_F Q$.
For a repulsive interaction ($\alpha_V>0$),  the pole and the branch cut are all in the lower half plane for any 
$Q \ll  \Lambda/v_F$ 
(for $Q >  \Lambda/v_F$, 
 \eq{eq:chiapp} is not valid).
For an attractive interaction ($\alpha_V<0$), the pole lies in the upper half plane for 
 $Q < Q^\ast$, where
$Q^\ast=\frac{2\Lambda}{v_F} e^{\frac{4\pi v_F}{\alpha_V}}$.
This represents modes that grow exponentially in time associated with the superconducting instability.
While all modes with center of mass momentum less than $Q^\ast$ become unstable,
the $Q=0$ mode exhibits
the fastest growth (the largest imaginary frequency in the upper half plane).

Now, let us understand the dynamics of the unstable modes in more detail.
To be concrete, we consider the following experimental set up.
A metal, whose ground state is a superconductor with a small gap, is initially kept from becoming superconductor by an external magnetic field.
At $t=0$, the magnetic field is turned off and at the same time a superconducting tip is 
brought close to the system for a short period of time.
The superconducting tip, which is a source of coherent Cooper pairs that can tunnel into the system, acts as a small pairing field $h_p$ applied to the system at $\vec r=0$ and $t=0$.
This will trigger an avalanche of unstable modes, driving the system to the superconducting ground state.
With other superconducting tips that are located at finite distances away from the initial superconducting tip, one can probe the spatial and temporal profile that arises from the real-time evolution of the condensate.
Within the period of the initial growth when the amplitude of the pair condensate is small,
the linear response theory is valid 
and
the condensate of momentum $\vec Q$ at time $t$ is written as
\bqa
P(\vec Q, t) = h_p~ 
\mathcal{G}_{pp}(\vec{Q},t), 
\eqa
where $h_p$ is the pairing field applied at $t=0$ and $\vec r=0$,
and
$\mathcal{G}_{pp}(\vec{Q},t)$
is the regarded Green's function that satisfies the boundary condition
$\mathcal{G}_{pp}(\vec{Q},t) = 0$ for $t<0$.

However,
$\chi^a_{pp}(\vec{Q}, \omega + i 0^+)$
 itself does not give the retarded Green's function because it has poles in the upper half plane.
The Fourier transformation of  
$\chi^a_{pp}(\vec{Q}, \omega + i 0^+)$
along the real axis of $\omega$ gives
\begin{eqnarray}
    G_{pp}(\vec{Q},t) =
    \Theta(-t) \mathcal{G}_{pp}^{(0)}(\vec{Q},t)
    +
    \Theta(t)
    \mathcal{G}_{pp}^{(1)}(\vec{Q},t),
    \label{eq:Gpp}
    \end{eqnarray}
    where
\begin{eqnarray}
\mathcal{G}_{pp}^{(0)}(\vec{Q},t) &=& -\frac{k_F}{\pi v_F}\Big(\frac{{4\pi v_F}}{\alpha_V}\Big)^2\Theta(Q^\ast-|\vec{Q}|) e^{\omega_{Q} t}\left[ e^{-\frac{4\pi v_F}{\alpha_V}}\frac{v_F^2Q^2}{4\Lambda}+ e^{\frac{4\pi v_F}{\alpha_V}}\Lambda\right], \nn
    \mathcal{G}_{pp}^{(1)}(\vec{Q},t) &=& 
    %\lim_{\Lambda_\omega \rightarrow \infty}\int^{\Lambda_\omega}_{-\Lambda_\omega} 
    \int 
    \frac{d\omega}{2\pi} e^{-i\omega t} ~
    \chi_{pp}^a(\vec{Q},\omega+i \eta).
\end{eqnarray}
$\mathcal{G}_{pp}^{(0)}(\vec{Q},t)$
is the contribution of the pole located in the upper half plane for $Q < Q^\ast$,
which can be picked up by extending the frequency integration of $\chi_{pp}^a(\vec{Q},\omega+i \eta)$ along the real axis with the infinite semi-circle in the upper half plane at $t<0$.
The hard cutoff $Q^\ast$ for the momentum integration reflects the fact that the Cooper pair mode is unstable only for $Q < Q^\ast$.
$\mathcal{G}_{pp}^{(1)}(\vec{Q},t)$
represents the contributions of the poles in the lower half plane.

\begin{figure}[h]
    \centering
    \includegraphics[width=.5\textwidth]{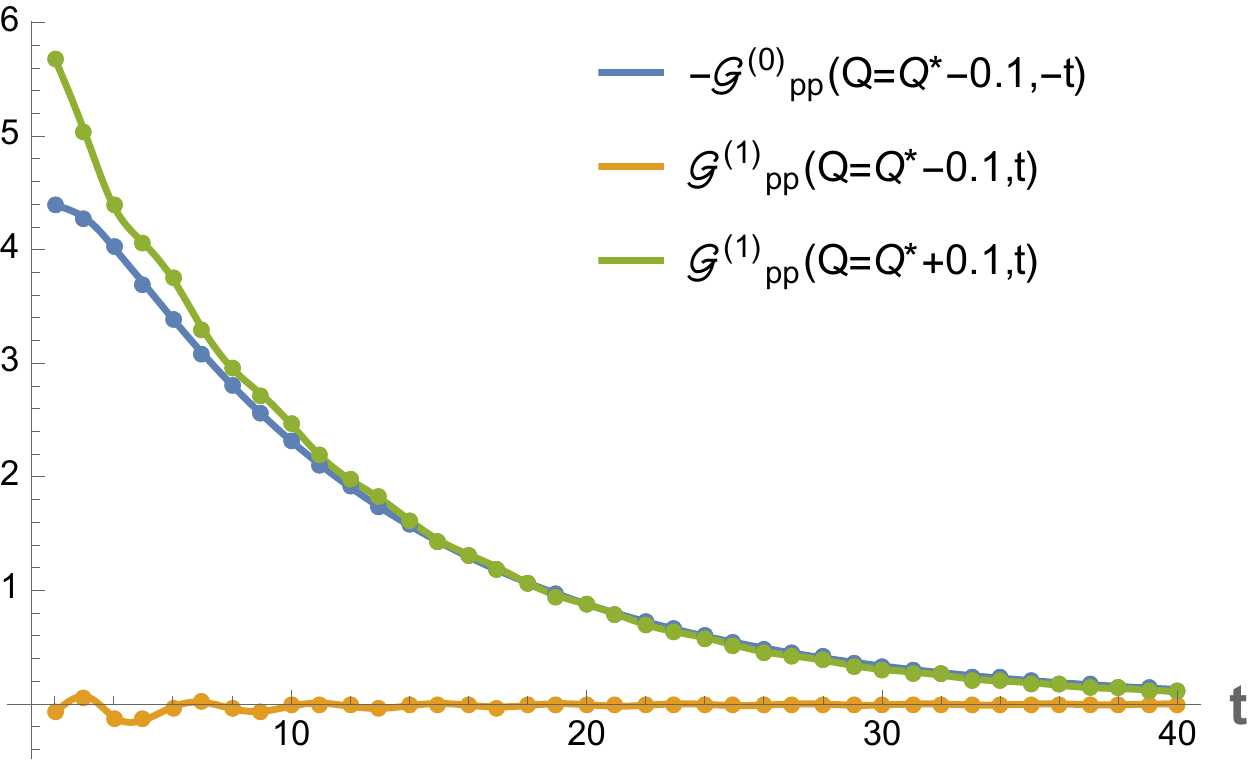}
    \caption{
    $-\mathcal{G}_{pp}^{(0)}(\vec{Q}^\ast-0.1,-t)$ 
    and
    $\mathcal{G}_{pp}^{(1)}(\vec{Q}^\ast \pm 0.1,t)$ 
    plotted as a function of $t > 0$ for
    $\eta=0.001$, $\Lambda=6$, $\Lambda_\omega=100$ and $\frac{\alpha_V}{4\pi v_F}=-0.5$.
    $\mathcal{G}_{pp}^{(1)}(\vec{Q},t)$ has been numerically evaluated with
    $m=\frac{k_F}{v_F}=\pi$ and $v_F=1$.
    }
    \label{fig:exponential_large_t}
\end{figure}

$G_{pp}(\vec{Q},t)$ is nonzero at all $t$,
and decays exponentially in the $|t| \rightarrow \infty$ limit because
$\mathcal{G}_{pp}^{(0)}(\vec{Q},t)$
and
$\mathcal{G}_{pp}^{(1)}(\vec{Q},t)$
decay at large negative and positive $t$, respectively,
as is shown in  Fig.~     \ref{fig:exponential_large_t}.
What \eq{eq:Gpp} describes is the evolution of a pair condensate that existed even before the pair field is applied at $t=0$,
where its amplitude gradually increases from zero to a finite value as $t$ increases from 
$-\infty$ to $0$.
While the condensate would have kept growing in $t>0$, the field  applied at $t=0$ alters the condensate into
$\mathcal{G}_{pp}^{(1)}(\vec{r},t)$ in $t>0$\footnote{
This is analogous to the situation where a ball slowly rolls down from the top of a mountain before it is kicked back so that it climbs back to the top in the $t \rightarrow \infty$ limit.
}.
To describe the situation in which the condensate amplitude is zero in $t<0$,
one has to add a time-dependent condensate that exists without the external source and cancels     \eq{eq:Gpp} in $t<0$.
Therefore, the retarded Green's function becomes
\begin{eqnarray}
    \mathcal{G}_{pp}(\vec{Q},t) =
    G_{pp}(\vec{Q},t) 
    -
    \mathcal{G}_{pp}^{(0)}(\vec{Q},t).
    \label{eq:retardedppG}
\end{eqnarray}
\eq{eq:retardedppG} satisfies the desired boundary condition,
$\mathcal{G}_{pp}(\vec{Q},t<0) =0$.
Furthermore,     
$\mathcal{G}_{pp}^{(0)}(\vec{Q},t)$ now captures the exponentially growing condensate in $t>0$.

\begin{figure}[h]
    \centering
    \includegraphics[width=.6\textwidth]{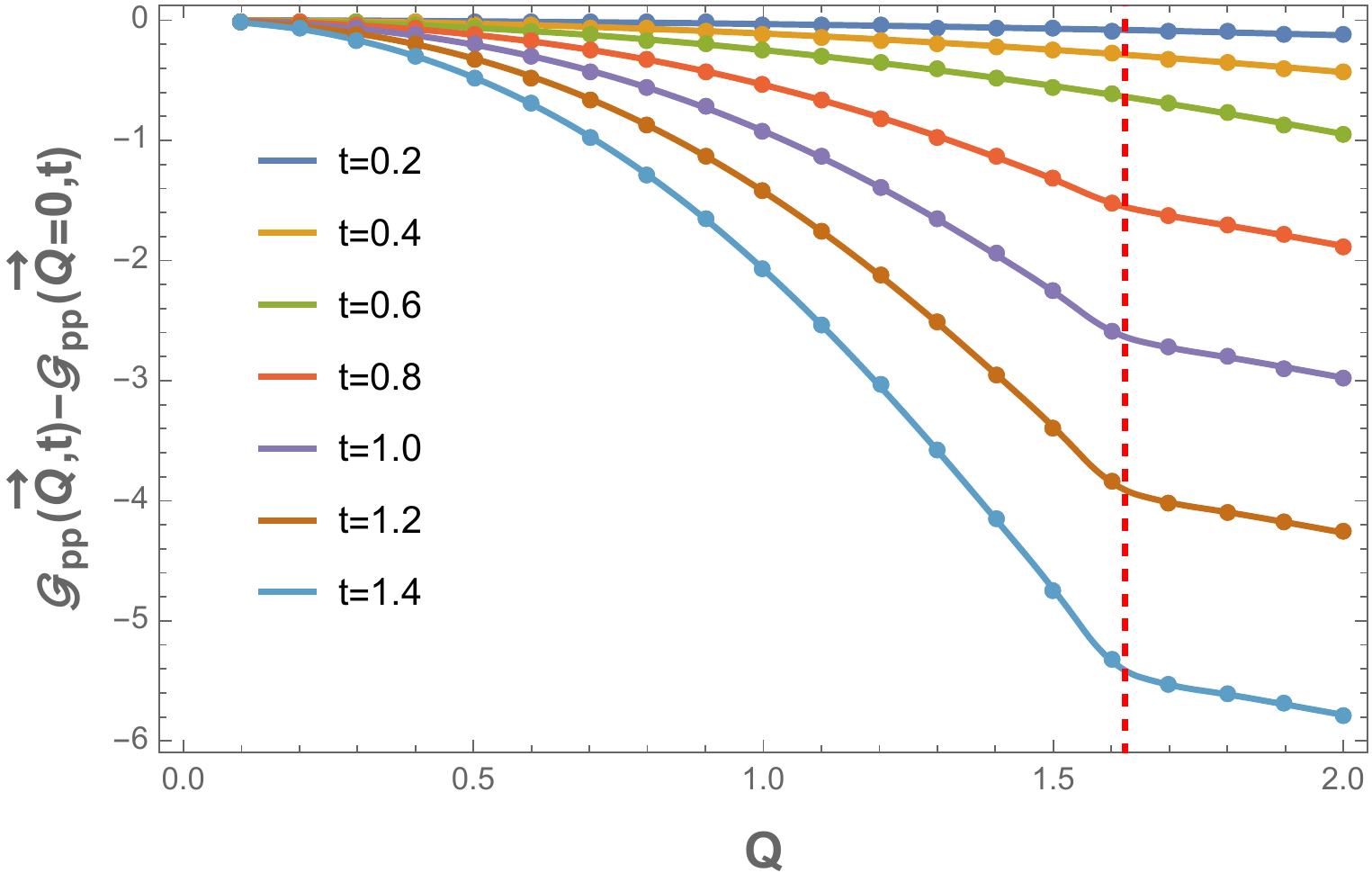}
    \caption{$\mathcal{G}_{pp}(\vec{Q},t)-\mathcal{G}_{pp}(0,t)$ plotted as a function of $Q$ for $0.2 \leq t \leq 1.4$
    with  $\eta=0.001$, $\Lambda=6$, $\frac{\alpha_V}{4\pi v_F}=-0.5$ and $\Lambda_\omega=100$. The red dashed line denotes the location of $Q^\ast$.
    }
    \label{fig:G_Q_t}
\end{figure}

We plot $\mathcal{G}_{pp}(\vec{Q},t)$ as a function of $Q$ 
in Fig.~\ref{fig:G_Q_t}. 
In the small $t$ limit, $\mathcal{G}_{pp}(Q,t)$ is independent of $Q$. 
This gives a $\delta$-function in real space, which describes localized pair condensate created by the local pair field applied at $t=0$.
With increasing $t$,
$\mathcal{G}_{pp}(Q,t)$ develops a non-trivial profile as condensates with different momenta grow at different rates. 
The mode with $Q=0$ grows at the fastest rate but all modes with $Q< Q^\ast$ grow independently as far as the amplitude of the condensate is small enough that the interaction among Cooper pairs can be ignored.
This gives rise to a spatial inhomogeneity in the phase of condensate.
To see this, 
we consider $t  \gg \Lambda/(v_F Q^\ast)^2$ but small enough that the amplitude of condensate is small.
In this case, 
$\mathcal{G}_{pp}^{(0)}(\vec{Q},t)$ gives the dominant contribution
and the Green's function is well approximated by
\begin{eqnarray}
    \mathcal{G}_{pp}(\vec{r},t)&\approx& -\mathcal{G}^{(0)}_{pp}(\vec{r},t) 
    =\frac{k_F}{\pi v_F} \frac{ v_F Q^\ast}{2\pi  }\left(\frac{4\pi v_F}{\alpha_V}\right)^2  \frac{ v_F^2 t (2+v_FQ^\ast t)- v_FQ^\ast r^2}{ 2v_F^2 t^3 }e^{-\frac{v_FQ^\ast r^2}{2v_F^2 t}+\frac{v_FQ^\ast}{2}t}.
    \label{eq:G_large_t}
\end{eqnarray}
\eq{eq:G_large_t} describe a diffusive behaviour of the exponentially growing pair condensate.
It is interesting to note that $\mathcal{G}_{pp}^{(0)}(\vec{r},t)$ is positive for $r<r_c$
with 
$r_c=\sqrt{\frac{v_F t(2+v_FQ^\ast t)}{Q^\ast}}$ 
while it becomes negative for  $r>r_c$.
At $r_{min}= \sqrt{\frac{v_F t(4+v_F Q^\ast t)}{Q^\ast}}$, the condensate 
 becomes most negative and its magnitude decreases algebraically in $t$ :
$\mathcal{G}_{pp}^{(0)}(\vec{r}_{min},t)=-\frac{k_F}{\pi v_F}\left(\frac{4\pi v_F}{\alpha_V}\right)^2\frac{v_F Q^\ast}{2\pi e^2 t^2}$. 
The inclusion of 
$\mathcal{G}_{pp}^{(1)}(\vec{r}_{min},t)$, which is exponentially small at large $t$, will modify the precise location of $r_c$, but won't remove the region of the condensate with phase difference $\pi$.
The appearance of the inverted condensate in $r>r_c$ is a consequence of the unstable modes with non-zero momenta.
The exponentially growing modes with non-zero $Q$ cause a destructive interference at $r_c$ and the inverted condensate in $r>r_c$.
With increasing time,  the in-phase condensate near $\vec r=0$ pushes  the inverted condensate to the region outside radius $r_c \sim  v_F t$ as the modes with non-zero $Q$ grows more slowly than the uniform condensate.
In the ultimate long-time limit, the interaction between Cooper pairs kicks in to stabilize the uniform superconducting state. 
However, the appearance of the transient superconducting condensate with phase shift $\pi$ is unavoidable in the initial time period when the amplitude of the condensate is still small.

Results discussed in this section can be in principle obtained by computing the density-density and pair-pair correlation functions directly from a microscopic model that include short-ranged interactions.
Then, what is the merit of using this low-energy effective theory?
First, our local effective field theory keeps all universal information and explains low-energy phenomena such as the collective modes in a self-contained manner.
This is in contrast to Landau's fixed point theory that can not describe the collective modes without including certain `high-energy' information which is not part of the fixed-point theory\footnote{
For example, in Landau's kinetic theory, 
the energy functional $E = \int dx d \theta d \theta' ~F(\theta, \theta') f(x,\theta) f(x,\theta')$ is used to describe the collective modes of Fermi surface, where
$f(x,\theta)$ describes the position dependent displacement of Fermi surface at angle $\theta$
and $F(\theta,\theta')$ is the Landau function.
However, this energy function actually includes the non-forward scattering that is independent of momentum transfer.}.
Second, the effective field theory makes the universal nature of physical predictions manifest.
The relations among low-energy observables that are determined from the fixed-point coupling functions are guaranteed to be universal because the coupling functions depend only on the universal low-energy data of the fixed point.
Finally, the present framework of local low-energy effective field theory can be readily generalized to non-Fermi liquids for which critical non-forward scatterings play even more important roles than in Fermi liquids\cite{BORGES2023169221}.

\section{Summary \label{sec:summary}}

In summary, we study Landau Fermi liquid and its instabilities within the low-energy effective field theory that is valid beyond the strict zero energy limit.
The local effective field theory should include general coupling functions that include non-forward scatterings and pairing interactions with non-zero center of mass momenta.
At low energies, 
the coupling functions
exhibit universal scaling behaviours when the momentum transfer and the center of mass momentum are comparable to the energy scale.
The scaling behaviour of the general coupling functions 
 determines various physical observables at low energies\cite{
PhysRevB.73.045128,
PhysRevB.69.121102,
2012LitJP..52..142P,
PhysRevB.86.155136}.
In particular, the dynamics of the low-energy collective modes is fully encoded within the momentum-dependent coupling functions.
This allows us to understand all low-energy physics of Fermi liquids within the low-energy effective field theory without resorting to microscopic theories.
We reproduce universal dynamics of the zero sound mode from the momentum-dependent coupling function in the particle-hole channel.
% \sungsik{
The coupling functions also contains the dynamical information about the unstable modes in the presence of instabilities. 
As an unstable normal Fermi liquid evolves toward the superconducting ground state,
we predict that it inevitably goes through a period of an inhomogenous superconductivity with a local phase inversion in the superconducting condensate 
due to the universal momentum-dependence of the renormalized coupling function
in the particle-particle channel. 
The local low-energy effective field theory valid away from the strict zero energy limit also reveals new types of instabilities of Fermi liquids.
Unlike the forward scattering amplitude that is exactly marginal, the non-forward scattering amplitudes are subject to non-trivial quantum corrections. 
If the strength of the bare attractive interaction exceeds a critical strength, it can drive instabilities toward symmetry broken states in particle-hole channels. 
The momentum-depedent coupling functions can be also tested 
more directly through the double photoemission spectroscopy\cite{ PhysRevLett.81.2148}.
% }

\section*{Acknowledgement}

We thank R. Shankar for 
 a discussion.
Research at the Perimeter Institute is supported in part by the
Government of Canada through Industry Canada, and by the Province of
Ontario through the Ministry of Research and Information.
SL also acknowledges the support by the Natural Sciences  and Engineering Research Council of Canada.

\bibliography{references}

\end{document}